\input harvmac.tex 
\input amssym.def
\input amssym
\input epsf.tex
\magnification\magstep1
\baselineskip 14pt
\parskip 6pt
\newdimen\itemindent \itemindent=32pt
\def\textindent#1{\parindent=\itemindent\let\par=\resetpar%
\indent\llap{#1\enspace}\ignorespaces}

\let\oldpar=\par
\def\resetpar{\oldpar\parindent=20pt\let\par=\oldpar}

\font\ninerm=cmr9 \font\ninesy=cmsy9
\font\eightrm=cmr8 \font\sixrm=cmr6
\font\eighti=cmmi8 \font\sixi=cmmi6
\font\eightsy=cmsy8 \font\sixsy=cmsy6
\font\eightbf=cmbx8 \font\sixbf=cmbx6
\font\eightit=cmti8
\def\eightpoint{\def\rm{\fam0\eightrm}
 \textfont0=\eightrm \scriptfont0=\sixrm \scriptscriptfont0=\fiverm
 \textfont1=\eighti  \scriptfont1=\sixi  \scriptscriptfont1=\fivei
 \textfont2=\eightsy \scriptfont2=\sixsy \scriptscriptfont2=\fivesy
 \textfont3=\tenex   \scriptfont3=\tenex \scriptscriptfont3=\tenex
 \textfont\itfam=\eightit  \def\it{\fam\itfam\eightit}%
 \textfont\bffam=\eightbf  \scriptfont\bffam=\sixbf
 \scriptscriptfont\bffam=\fivebf  \def\bf{\fam\bffam\eightbf}%
 \normalbaselineskip=9pt
 \setbox\strutbox=\hbox{\vrule height7pt depth2pt width0pt}%
 \let\big=\eightbig  \normalbaselines\rm}
\catcode`@=11 %
\def\eightbig#1{{\hbox{$\textfont0=\ninerm\textfont2=\ninesy
 \left#1\vbox to6.5pt{}\right.\n@@space$}}}
\def\vfootnote#1{\insert\footins\bgroup\eightpoint
 \interlinepenalty=\interfootnotelinepenalty
 \splittopskip=\ht\strutbox %
 \splitmaxdepth=\dp\strutbox %
 \leftskip=0pt \rightskip=0pt \spaceskip=0pt \xspaceskip=0pt
 \textindent{#1}\footstrut\futurelet\next\fo@t}
\catcode`@=12 %
\def \de{\delta}
\def \De{\Delta}
\def \si{\sigma}

\def \pr{\partial}
\def \tr{{\rm tr }}

\def \tz{{\tilde z}}
\def \hI{{\hat{\cal I}}}

\def \J{{\rm J}}

\def \r{\big \rangle}

\def \vep{\varepsilon}
\def \half{{\textstyle {1 \over 2}}}

\def \quar{{\textstyle {1 \over 4}}}
\def \ts{\textstyle}

\def \d{{\rm d}}
\def \v{{\rm v}}
\def \x{{\rm x}}
\def \y{{\rm y}}

\def \ty{{\tilde {\rm y}}}

\def \A{{\cal A}}
\def \B{{\cal B}}
\def \C{{\cal C}}
\def \D{{\cal D}}
\def \E{{\cal E}}
\def \F{{\cal F}}
\def \G{{\cal G}}
\def \H{{\cal H}}
\def \I{{\cal I}}
\def \J{{\cal J}}

\def \L{{\cal L}}
\def \M{{\cal M}}
\def \N{{\cal N}}
\def \O{{\cal O}}
\def \P{{\cal P}}
\def \Q{{\cal Q}}
\def \R{{\cal R}}
\def \S{{\cal S}}

\def \V{{\cal V}}

\def \DD{{\sl D}}

\def \dal{{\dot \alpha}}
\def \dbe{{\dot \beta}}
\def \dga{{\dot \gamma}}
\def \bQ{{\bar Q}}
\def \bS{{\bar S}}
\def \bsi{\bar \sigma}
\def \bj{{\bar \jmath}}
\def \bJ{{\bar J}}

\def \ty{{\tilde {\rm y}}}

\def \x{{\rm x}}
\def \y{{\rm y}}
\def \t{{\rm t}}
\def \z{{\rm z}}

\def \vphi{{\varphi}}

\def \dal{{\dot \alpha}}
\def \dbe{{\dot \beta}}
\def \dga{{\dot \gamma}}

\def \bsi{\bar \sigma}

\def \bvphi{{\overline \vphi}}
\def \blam{{\overline \lambda}}
\def \bpsi{{\overline \psi}}
\def \bmu{{\overline \mu}}
\def \ua{{\underline a}}
\def \ub{{\underline b}}

\def \bx{\bar x}

\def \al{{\alpha}}
\def \tr{{\rm tr}}
\def \u{{\rm u}}
\def \be{\beta}

\def \by{\bar x}

\def \bJ{{\bar J}}
\def \bQ{\bar Q}
\def \bS{\bar S}
\def\rmY{{\rm u}}
\def\hu{{\hat {\rm u}}}
\def \wtN{\tilde N}
\def \ttr{\widetilde{\tr}}
\def\toinf#1{\mathrel{\mathop{\longrightarrow}\limits_{\scriptstyle{#1}}}}

\font \bigbf=cmbx10 scaled \magstep1


\lref\fadho{F.A. Dolan and H. Osborn, {\it On short and semi-short
representations for four-dimensional superconformal symmetry},
Ann. Phys. 307 (2003) 41, hep-th/0209056.}

\lref\char{M. Bianchi, F.A. Dolan, P.J. Heslop and H. Osborn,
{\it $\N=4$ Superconformal Characters and Partition Functions},
Nucl. Phys. B767 (2007) 163, hep-th/0609179.}

\lref\mald{J.~Kinney, J.M.~Maldacena, S.~Minwalla and S.~Raju,
{\it An Index for $4$ Dimensional Super Conformal Theories},
Comm. Math. Phys. 275 (2007) 209, hep-th/0510251.}

\lref\janik{R.A. Janik and M. Trzetrzelew, { \it  Supergravitons from one loop 
perturbative $N=4$ SYM}, arXiv:0712.2714 [hep-th].}

\lref\andrews{G.E. Andrews, R. Askey and R. Roy, {\it Special Functions},
Cambridge University Press, Cambridge, 1999.}

\lref\basic{G. Gasper and M. Rahman, {\it Basic Hypergeometric Series},
Cambridge University Press, Cambridge, 2004.}

\lref\bofengt{B. Feng, A. Hanany, Y.H. He,
 {\it Counting Gauge Invariants: The Plethystic Program}, 
 JHEP (2007)  0703:090, hep-th/0701063.}

\lref\nak{Y. Nakayama, 
{\it Finite $N$ Index and Angular Momentum Bound from Gravity}, 
hep-th/0701208.}

\lref\romelf{C. R\"omelsberger,
 {\it Counting chiral primaries in $\N = 1$, $d=4$ superconformal field theories},
Nucl. Phys. B747 (2006) 329, hep-th/0510060.}

\lref\romel{C. R\"omelsberger,
{\it Calculating the Superconformal Index and Seiberg Duality},
arXiv:0707.3702 [hep-th].}

\lref\FN{F.A. Dolan,
{\it Counting BPS operators in $\N=4$ SYM}, Nucl. Phys. B790 (2008) 432,
arXiv:0704.1038 [hep-th].}

\lref\spi{V.P. Spiridonov, {\it On the Elliptic Beta Function}, Russian
Math. Surveys (2001) 56, 185\semi
{\it An Elliptic Beta Integral}, in  Proceedings of the Fifth International 
Conference on Difference Equations, ed. S. Elaydi, Taylor \& Francis, London, 2002.}

\lref\spit{V.P. Spiridonov. {\it Classical Elliptic Hypergeometric Functions and
Their Applications}, Rokko Lectures in Mathematics, vol.18, 
Elliptic Integrable Systems, ed. M. Noumi and K. Takasaki (2005),arXiv:math/0511579 [math.CA].} 

\lref\rains{E.M. Rains, {\it Transformations of Elliptic Hypergeometric
Integrals}, Ann. Math., to appear, math/0309252.}

\lref\rainsB{E.M. Rains, private communication.}

\lref\proof{V.P. Spiridonov, {\it Short proofs of elliptic beta integrals},
Ramanujan Journal 13 (2007) 265, math/0408369 [math.CA]\semi
E.M. Rains and V.P. Spiridonov, {\it  Determinants of elliptic hypergeometric 
integrals}, arXiv:0712.4253 [math.CA].}

\lref\Nas{B. Nassrallah and M. Rahman, {\it Projection formulas, a reproducing
kernel and a generating function for $q$-Wilson polynomials}, SIAM J. Math. Anal.
16 (1985) 186\semi
M. Rahman, {\it An integral representation of a ${}_{10}\vphi_9$ and
continuous biorthogonal ${}_{10}\vphi_9$ rational functions}, Can. J. Math.
38 (1986) 605.}

\lref\askey{R. Askey, {\it Beta Integrals in Ramanujan's Papers}, Ramanujan
Revisited, Academic Press, London 1988, pp. 561-590.}

\lref\Kut{D. Kutasov, {\it A Comment on Duality in $N=1$ Supersymmetric
Non-Abelian Gauge Theories}, Phys. Lett. B 351 (1995) 230, hep-th/9503086.}

\lref\KS{D. Kutasov and A. Schwimmer, {\it On Duality in Supersymmetric
Yang-Mills Theory}, Phys. Lett. B 354 (1995) 315, hep-th/9505004.}

\lref\KSS{D. Kutasov, A. Schwimmer and N. Seiberg, {\it Chiral Rings,
Singularity Theory and Electric-Magnetic Duality},
Nucl. Phys. B459 (1996) 455, hep-th/9510222.}

\lref\intrilo{K.A. Intriligator,
{\it New RG Fixed Points And Duality In Supersymmetric $Sp(N_c)$ And
$SO(N_c)$ Gauge Theories},
Nucl. Phys. B448 (1995) 187, hep-th/9505051.}

\lref\IntS{K. Intriligator and N. Seiberg, {\it Duality, Monopoles, Dyons,
Confinement and Oblique Confinement in Supersymmetric $SO(N_c)$ Gauge Theories},
Nucl. Phys. B444 (1995) 125, hep-th/9503179.}

\lref\Sei{N. Seiberg, {\it Electric-Magnetic Duality in Supersymmetric
Non-Abelian Gauge Theories}, Nucl. Phys. B435 (1995) 129, hep-th/9411149.}

\lref\Except{J. Distler and A. Karch, {\it N=1 Dualities for Exceptional Gauge Groups
and Quantum Global Symmetries}, Fortsch. Phys. 45 (1997) 517, hep-th/9611088\semi
A. Karch, {\it More on N=1 Dualities and Exceptional Gauge Groups}, Phys. Lett.
B405 (1997) 280, hep-th/9702179.} 

\lref\intrilp{K.A. Intriligator and P. Pouliot,
{\it Exact Superpotentials, Quantum Vacua and Duality in Supersymmetric
$Sp(N_c)$ Gauge Theories},
Phys. Lett.  B353 (1995) 471, hep-th/9505006.}

\lref\Tern{J. Terning, {\it Modern Supersymmetry}, Clarendon Press, Oxford 2006.}

\lref\mep{F.A. Dolan, {\it Character Formulae and Partition Functions in
Higher Dimensional Conformal Field Theory},
 J. Math. Phys. 47 (2006) 062303, hep-th/0508031.}

{\nopagenumbers
\rightline{DAMTP/08-07}
\rightline{DIAS-STP-08-02}
\rightline{SHEP-08-06}
\rightline{arXiv:0801.4947 [hep-th]}
\vskip 1.5truecm
\centerline {\bigbf Applications  of the Superconformal Index for Protected Operators}
\vskip 3pt
\centerline {\bigbf   and $q$-Hypergeometric Identities to $\N=1$ Dual Theories}
\vskip  6pt
\vskip 2.0 true cm
\centerline {F.A. Dolan${}^{\ast,\dagger}$
and H. Osborn${}^\ddagger$}

\vskip 12pt
\centerline {${}^\ast$Institi\'uid Ard-l\'einn Bhaile \'Atha Cliath,}
\centerline {(Dublin Institute for Advanced Studies,)}
\centerline {10 Burlington Rd., Dublin 4, Ireland}
\vskip  6pt
\centerline {${}^\dagger$School of Physics and Astronomy, Southampton University,}
\centerline {Highfield, Southampton SO17 1BJ, England}
\vskip  6pt
\centerline {${}^\ddagger$Department of
Applied Mathematics and Theoretical Physics,}
\centerline {Wilberforce Road, Cambridge CB3 0WA, England}
\vskip 1.5 true cm

{\eightpoint
\parindent 1.5cm{

{\narrower\smallskip\parindent 0pt

The results of R\"omelsberger for a $\N=1$ superconformal index counting protected
operators, satisfying a BPS condition and which cannot be combined to form long 
multiplets, are analysed further. The index is expressible in terms of single
particle superconformal characters  for $\N=1$ scalar and vector multiplets. 
For SQCD, involving $SU(N_c)$ gauge groups and appropriate numbers of flavours $N_f$,
the formula used to construct the index may be proved to give identical results for 
theories linked by Seiberg duality using recently proved theorems for $q$-series 
elliptic hypergeometric integrals. The discussion is also extended to Kutasov-Schwimmer 
dual theories in the large $N_c, N_f$ limit and to dual theories with $Sp(N)$
and $SO(N)$ gauge groups. For the former, a transformation identity for elliptic
hypergeometric integrals directly verifies that the index is the same for the 
electric and magnetic theories. For $SO(N)$ theories the corresponding result may also
be obtained from the same basic identity. An expansion of the index to several orders 
is also obtained in a form where the detailed protected operator content may be read off.
Relevant mathematical results are reviewed.


\narrower}}

\vfill
\line{\hskip0.2cm E-mails:
{{\tt
fdolan@stp.dias.ie,
H.Osborn@damtp.cam.ac.uk}}\hfill}
}

\eject}

\pageno=1

\noindent
\newsec{Introduction}
\noindent
A remarkable new insight into the dynamics of supersymmetric quantum field
theories was the discovery by Seiberg in the 1990's of dualities 
analogous to those in soluble two dimensional integrable models \Sei, 
for a textbook discussion see \Tern. For a $\N=1$ gauge theory with
gauge group $G$ and a suitable number $N_f$ of chiral matter `quark' fields, 
belonging the fundamental representation of $G$ and transforming under a flavour
symmetry group $F$, there is a duality between the initial 
`electric' theory and an associated `magnetic' theory with a dual gauge group
$\tilde G$ but the same flavour symmetry $F$. In the dual magnetic theory,
besides the appropriate `quark' fields, the matter fields also include chiral
`mesons' to match with the corresponding electric theory.
Both electric and magnetic theories are asymptotically free 
but they have a common IR fixed point realising a non-trivial interacting 
$\N=1$ superconformal theory. As usual in dual theories the strong coupling regime 
of the electric theory corresponds to the weak coupling regime of the magnetic one, 
and vice-versa. In the
canonical example $G=SU(N_c)$ and $F=SU(N_f)\times SU(N_f) \times U(1)_B \times
U(1)_R$ and with ${3\over 2} N_f \le N_c \le 3N_f$ then $\tilde G = SU(N_f -N_c)$.
Each conjectured duality is justified by many non-trivial consistency checks.
The original Seiberg dualities have also been extended to different
gauge groups \intrilo\ and theories with further fields \refs{\Kut,\KS,\KSS} showing 
the existence of a plethora of superconformal IR fixed points in $\N=1$ 
supersymmetric field theories linked by RG flows after introducing mass terms or 
other relevant perturbations.
 
More recently the detailed operator content of four dimensional 
superconformal gauge theories has been intensively investigated. A critical
issue is to distinguish between protected operators satisfying a BPS condition
and whose scale dimensions $\Delta$ saturate an associated unitarity bound and those 
operators which are not so constrained with a scale dimension determined by the
detailed dynamics. In $\N=4$ theories the former belong
to short or semi-short supermultiplets while the latter form long multiplets
with $\Delta$ depending on $g$ the coupling so that they may disappear from the 
spectrum in the strong coupling limit. Since semi-short multiplets may
combine to form long multiplets which gain anomalous dimensions in perturbation
theory the counting of protected operators, satisfying BPS constraints, is a 
not an immediately straightforward issue. In \mald\ Kinney {\it et al} formulated 
an index for general $\N$ superconformal theories such that contributions from any
combinations of multiplets forming a long multiplet cancel and hence only
protected operators are relevant. The index is then a topological invariant under
smooth deformations preserving superconformality
and was calculated in \mald\ to give the same results for $\N=4$ theories both at
weak coupling and also at strong coupling through the AdS/CFT correspondence, see 
also \janik. The index in various sectors may also be obtained \char\ by considering
suitable limits of partition functions for counting gauge singlet operators where 
the relevant characters involve the supertrace, or equivalently contain a factor 
$(-1)^F$, and the limit ensures no long multiplet contribution. These results 
were applied also in \char\
to discuss $\N=4$ theories with an $SU(N)$ gauge group in the large $N$ limit.

For $\N=1$ theories the basic contributions to the index are expressible as
$SU(2,1)$ characters. For such theories  R\"omelsberger \refs{\romelf,\romel}\ also 
constructed an index which is essentially equivalent to that of \mald\ in this case.
R\"omelsberger further gave a prescription for determining the index at the
non trivial IR fixed points related by Seiberg duality and then showed
that there was a very non-trivial matching of the two independent
electric and magnetic expressions for the index by considering a series
expansion up to a certain order in particular cases. In general to calculate the 
index it is necessary to identify a supercharge $\Q$, with associated 
adjoint\foot{The adjoint here is defined, for a space of states formed by local
field operators $\phi$ acting on $|0\rangle$, by a scalar product determined by
the two point functions for $\phi$. It differs from the usual conjugation
so that, for any operator $\O$, $\O^+ = U^{-1}\O^\dagger U$ for $U^\dagger =U$,
\char. Thus for the dilation operator $H^+=H$ although $H^\dagger = - H$.} $\Q^+$, 
such that
\eqn\QQH{
\{ \Q , \Q^+ \} = 2\H \, , \qquad \Q^2 = 0 \, ,
}
so that $\H$ has a positive semi-definite spectrum. The index is then formed by
the supertrace for states belonging to the kernel of $\H$
and so belonging to the cohomology of  $\Q,\Q^+$. The generators commuting with 
$\Q,\Q^+$ in $\N=1$ theories are then
\eqn\suto{
\M_A{}^{\! B} = \pmatrix{ M_\alpha{}^{\! \beta} + \half \de_\alpha{}^{\! \beta} \R & 
\P_{\alpha} \cr \noalign{\vskip 2pt}
- {\bar \P}{}^{\beta} & -\R \cr} \, , \qquad  {\bar \P}{}^{\beta} = (\P_\beta)^+ \, ,
}
which satisfy the Lie algebra for $SU(2,1)$, $[ \M_A{}^{\! B}, \M_C{}^{\! D} ] = 
\de_C{}^{\! B} \M_A{}^{\! D} - \de_A{}^{\! D} \M_C{}^{\! B}$. In \suto\ 
$M_\alpha{}^{\! \beta} = ( M_\beta{}^{\alpha})^+$ contains the generators
$J_3,J_\pm$ for the $SU(2)$ subgroup acting on chiral spinors while
\eqn\RR{
\R = R + 2{\bar J}_3 + c \H \, ,
}
with $R$ the generator for $U(1)_R$. The index may then be defined by
\eqn\indt{
I(t,x) = \tr_{{\rm ker} \H} \big ( (-1)^F t^\R x^{2J_3} \big ) \, ,
}
although this may be extended by further variables related to additional
symmetries.

In the prescription of R\"omelsberger \romel\ for $\N=1$ superconformal theories
the index is first determined on `single particle states' giving
\eqn\ione{\eqalign{
i(t,x,h,g) ={}& {2t^2 - t(x+x^{-1})\over (1-tx)(1-tx^{-1})} \, \chi_{\rm adj.}(g)\cr
&{} + \sum_i {t^{r_i}\chi_{R_F,i}(h)\chi_{R_G,i}(g) -
t^{2-r_i}\chi_{{\bar R}_F,i}(h)\chi_{{\bar R}_G,i}(g) \over
(1-tx)(1-tx^{-1})}\, , }
}
which depends also on the symmetry group elements $g\in G, \, h \in F$.
In \ione\ the first term represents the contribution for gauge fields belonging
to the adjoint representation of $G$ and the sum corresponds to chiral matter fields 
$\varphi_i$ transforming under  gauge group representations $R_{G,i}$,
a flavour symmetry representations $R_{F,i}$, with  $\chi_{R_F,i}(h),\chi_{R_G,i}(g)$ 
the appropriate characters. The terms proportional to $t^{r_i}$ and $t^{2-r_i}$
result from a chiral scalar with $R$-charge $r_i$ and the fermion
descendant, with ${\bar \jmath} = {1\over 2}$, of the conjugate anti-chiral partner 
with $R$-charge $-r_i$. In order to 
determine the index for all gauge singlet operators, as relevant for confining
theories, this is then inserted into the `plethystic' exponential \bofengt\ giving
\eqn\itwo{
I(t,x,h) = \int_G \! \d \mu(g) \, \exp \bigg ( \sum_{n=1}^\infty {1\over n} \,
i \big (t^n ,x^n, h^n , g^ n\big ) \bigg ) \, ,
}
for $\d \mu(g)$ the $G$ invariant measure.
A unitary superconformal representation would require in \ione\ $r_i\ge {2\over 3}$ with
$r_i = {2\over 3}$ corresponding to a free field. In confining theories for chiral scalars 
belonging to non trivial representations of the gauge group this may be relaxed although
it is necessary here that $r_i+r_j \ge {2\over 3}$ if $R_{G,i} \times R_{G,j}$
contains the identity representation and there is a corresponding composite gauge singlet
$\varphi_i \cdot \varphi_j$, unless this operator is coupled to a dynamical field in the 
superpotential and so is constrained by equations of motion.
In general we assume here unitary positive energy
representations of $SU(2,1)$ requiring therefore $0<r_i<1$.

The interpretation of $I$ as a superconformal index requires that the result for 
\itwo\ should have an expansion of the form
\eqn\expI{
I(t,x,h) = \sum_{q,j,R_F} n_{q,j,R_F} \, 
{t^q \, \chi_{2j+1}(x) \over (1-tx)(1-tx^{-1})} \, \chi_{R_F}(h) \, ,
}
where $\chi_{2j+1}$ are $SU(2)$ characters, and with  $n_{q,j,R_F}$ integer 
coefficients which determine the spectrum of protected operators in the $\N=1$ 
superconformal theory. Contributions to the sum in \expI\ for different
supermultiplets are found in appendix A. Long multiplets are absent
but contributions are present for chiral operators when $q=r$, the $R$-charge,
with sign $(-1)^{2j}$ but there may also be contributions for other protected 
operators when $q = 2 +2\bj+r$ and for sign $-(-1)^{2j+2\bj}$.

Despite generating formulae for the index which are in impressive agreement
for dual superconformal theories the status of the results for the $\N=1$ 
superconformal index given by \ione\ and \itwo\  is nevertheless not immediately 
clear, even for theories with no superpotential. Unlike the discussion in \mald\ 
for the $\N=4$ case there is no continuous link between the free case and the 
strong coupling limit, which is relevant for an IR fixed point, while preserving 
superconformal symmetry so that the index is well defined. The index formula in 
the asymptotically free limit gives different results since then 
$r_i = {2\over 3}$ for all $\varphi_i$.

Nevertheless we explore the consequences of the formulae for the index given by \ione\ 
and \itwo\ in a significant number of examples and verify in many cases
that the same result is obtained for both the electric and magnetic
theories linked by Seiberg duality and its extensions, and hence develop the tests in  
\romel\ further. In general this requires non trivial identities for the 
group integrals in \itwo\ for $G$ and its dual $\tilde G$ which are then equivalent to
identities for $q$-hypergeometric elliptic integrals. In some cases the magnetic
theory is such that the dual gauge group $\tilde G$ is trivial. The expression for the 
magnetic index then requires no group integration so that showing the index identity 
requires the evaluation of the integral defining the index in the electric theory. 

A particular example arises for $N_c=2, \, N_f=3$, which is perhaps the simplest
non trivial case. The electric theory defines a contour integral in one variable
while the magnetic theory provides an explicit evaluation. However, verifying this
is very non trivial, a special case is related to a result 
found by Nassrallah and Rahman for an extension of the usual beta integral \Nas. 
A generalisation of the Nassrallah-Rahman theorem by Spiridonov \spi,
involving elliptic gamma functions, is shown here to be directly
equivalent to the required $N_c=2, \, N_f=3$ superconformal index identity.
This provides an important clue as to the appropriate mathematical context
for showing how the electric and magnetic indices are equal in more general cases.
Identities obtained by Rains \rains\ linking multi-dimensional $q$-hypergeometric 
integrals, which reduce to the results of Spiridonov in special cases,  are sufficient 
to prove compatibility of the formulae for indices obtained by applying \ione\ and
\itwo\ with Seiberg duality in a wide range of cases.

The applicability of these results depends crucially on the detailed form of
\ione\ and \itwo. For the chiral matter fields a general term in \ione\ has the form
\eqn\imat{
i_S(p,q,y) = {t^{r}z -t^{2-r} z^{-1}\over (1-tx)(1-tx^{-1})} =
{y - pq/y \over (1-p)(1-q)} \, , \quad p=tx, \ q = tx^{-1} \, , \ y=t^r z \, ,
}
and then in \itwo
\eqn\eGam{
\Gamma(y;p,q) = \exp \bigg ( \sum_{n=1}^\infty {1\over n} \,
i_S(p^n,q^n,y^n ) \bigg ) = \prod_{j,k\ge 0}
{1-y^{-1} p^{j+1}q^{k+1}\over 1 - y\,  p^j\, q^k} \, ,
}
where $\Gamma(y;p,q)$ is an elliptic Gamma function and we  assume $p,q$ real
and $0\le p,q<1$.
Furthermore for the gauge field part of \ione\ we may define
\eqn\igauge{
i_V(p,q)= {2t^2 - t(x+x^{-1})\over (1-tx)(1-tx^{-1})} = -{p\over 1-p}- {q\over 1-q}
= 1 - {1-pq \over (1-p)(1-q)} \, ,
}
and then apply
\eqn\eThe{\eqalign{
\exp \bigg ( \sum_{n=1}^\infty {1\over n} \, i_V(p^n, q^n) ( z^n + z^{-n} ) \bigg ) 
= {}& {\theta(z;p)\, \theta(z;q) \over (1-z)^2} \cr
\noalign{\vskip -4pt}
= {}& {1\over (1-z)(1-z^{-1})\, \Gamma(z;p,q)\Gamma(z^{-1};p,q)}\, , \cr
\exp \bigg ( \sum_{n=1}^\infty {1\over n} \, i_V(p^n, q^n) \bigg ) 
= {}& (p;p)\, (q;q) \, , \cr}
}
where the theta  function and $(p;p)$ are infinite products  defined by
\eqn\defth{
\theta(z;p) =  {\ts \prod_{j\ge 0}} (1-zp^j)(1-z^{-1}p^{j+1} ) \, , \quad
(x;p) =  {\ts \prod_{j\ge 0}} (1-xp^j) \, .
}

The detailed discussion in this paper is as follows.  In section 2 the
superconformal transformation properties of $\N=1$ chiral scalar and 
vector multiplets are described. For free theories it is shown how
expressions for the index are constructed which are in accord with
the results \ione\ and \itwo\ given above but with the $R$-charge restricted
to its free field value.
In section 3, the dual Seiberg and Kutasov-Schwimmer theories, with $SU(N_c)$
gauge groups and $SU(N_f) \times SU(N_f)$ flavour symmetry, are reviewed and 
the single particle indices are obtained by applying \ione.
The multi-particle indices for these theories which are given by \itwo\ are then 
shown to agree in a certain large $N_c$, $N_f$ limit in section 4.  The case
of Seiberg duality for $(N_c,N_f)=(2,3)$ is discussed in detail
in section 5.  Section 6 extends to the general $(N_c,N_f)$ case where a theorem due to
Rains is shown to demonstrate that the results for the index in the electric
and magnetic theories are identical. 
Section 7 consider dual theories with $Sp(2N)$ gauge groups. With similar constructions
the index is shown to agree for both theories as a consequence of a related theorem.
As in section 6 the final result depends on non trivial integral identities. We also
discuss in section 8 dual theories with $SO(N)$ gauge groups where the chiral matter fields
belong to the vector representation. The resulting elliptic hypergeometric integrals are 
similar in form to the previous cases and the required identities can be found by expressing
them in terms of the corresponding integrals for the $Sp(2N)$ and using the
associated identity proven by Rains.
We also consider an expansion in one simple case and verify that the result is in
accord with \expI\ to the order calculated. 

Various appendices with miscellaneous mathematical details are included.  
Appendix A gives a discussion of
$\N=1$ superconformal representation theory and derives expressions
for the characters for different representations. The limits which are 
appropriate for the index and which are relevant for section 2 are also discussed.
In appendix B we summarise some general results for group characters which are
used in the main text while in appendix C we show how some corrections to the 
large $N$ limit discussed in section 4 can be calculated. 
Appendix D describes some properties of the essential elliptic Gamma functions 
introduced in \eGam\ and \defth. Identities given here are used in appendix E to 
outline how the  single variable elliptic hypergeometric integral,
that gives the index in the  simple example  for the electric theory 
when $N_c=2, \, N_f=3$, may be evaluated 
in agreement with the result determined by the corresponding magnetic theory.
Although a special case, the methods used in this calculation are illustrative of those
necessary to obtain more general results.

\newsec{$\N=1$ Superconformal Transformations and Chiral Fields}

The $\N=1$ superconformal algebra contains besides the usual supercharges,
$ Q_{\alpha}, \bQ_{\smash {\dal}}$, $\{Q_\al,\bQ_{\smash {\dal}}\}=
2P_{\al\dal}$, also their conformal partners, $S^\alpha,  \bS{}^{\dal}$, 
$\{\bS{}^{\dal},S^\alpha\} = 2K^{\dal\alpha}$, the generator of special
conformal transformations. For a superconformal primary field $\O$ then
$|\O\rangle = \O(0)|0\rangle$ is annihilated by 
$S^\alpha,  \bS{}^{\dal}$ and forms a lowest weight state for a supermultiplet.
The state
has scale dimension $\Delta$ and $R$-charge $r$ if $[ H , \O(0) ] = \Delta \O(0)$,
$[R, \O ] = r \O$, and the supermultiplet then has a basis formed by the action
of $ Q_{\alpha}, \bQ_{\smash {\dal}}, P_{\al\dal}$ on $|\O\rangle$. A chiral
field is such that $\bQ_{\smash {\dal}} |\O\rangle = 0$. As a consequence of
$\{ \bS{}^{\dal} , \bQ_{\smash {\dbe}}\}  |\O\rangle = 0$ the scale dimension is
then determined by the $R$-charge
\eqn\DR{
\Delta = {\ts{3\over 2}}r \, ,
}
and $\O$ must belong only  to a $(j,0)$ spin representation.

For a chiral scalar field $\vphi$ the action of the chiral supercharges
$Q_\alpha, S^\alpha$ is then
\eqn\QSfree{\eqalign{
\big [ Q_{\alpha} , \vphi \big ]  = {}& \psi_\alpha \, , \qquad
\big \{ Q{}_{\alpha} , \psi _{\beta} \big \} =
\vep_{\alpha\beta} F \, , \qquad \big [ Q_{\alpha} , F \big ]  = 0 \, , \cr
\big \{ S^{\beta} , \psi_{\al} \big \} = {}& 6r\, \de_{\al}{}^{\!\be} \vphi \, , \qquad
\big [  S^{\alpha} , F \big ] = - 2(3r-2) \, \vep^{\al\be} \psi_\beta\, , }
}
where the $S$ action is determined by consistency with the superconformal algebra.
Furthermore for $\bQ_{\smash {\dal}}$ the algebra also requires
\eqn\QbarS{
\big \{ \bQ_{\smash {\dal}} , \psi_\alpha \big \} = 2i \,\pr_{\al\dal}
\vphi \, , \qquad \big [ \bQ_{\smash {\dal}}  , F \big ] =  2 i \, \vep^{\be\al}
\pr_{\al \dal} \psi_{\be} \, .
}
For a chiral $(\half,0)$ spinor field $\lambda_\alpha$ we have similarly
\eqn\QVfree{\eqalign{
\big \{ Q_{\alpha} , \lambda_\beta \big \}  = {}& f_{\alpha\beta} +
\vep_{\alpha\beta}\, i\DD  \, , \qquad\qquad\big [ Q{}_{\alpha} , f_{\beta\gamma} \big ] 
= \vep_{\alpha\beta}\, \mu_\gamma + \vep_{\alpha\gamma}\, \mu_\beta \, , \cr
\big [ Q_{\alpha} , \DD \big ]  =  {}& i\, \mu_\alpha \, , \qquad\qquad \qquad\qquad
\big \{ Q_{\alpha} , \mu_\beta \big \}  = 0   \, , \cr
\big [ S^{\gamma} , f_{\al\be} \big ] = {}& 2(3r+1) \, 
\de_{(\al}{\!}^{\gamma} \lambda_{\be)}\, , \qquad  \ \,
\big [  S^{\beta} , \DD \big ] = 3(r-1)i \, \vep^{\be\al}\lambda_\al \, , \cr
\big \{ S^\beta , \mu_\alpha \big \} = {}& 
- 3(r -1)\, \vep^{\beta\gamma}f_{\alpha\gamma} - (3r +1)i\, \delta_\al{\!}^\be \DD \, , }
}
with $f_{\alpha\beta}= f_{\beta\alpha}$, and
\eqn\QbarV{\eqalign{
\big [ \bQ_{\smash {\dal}} , f_{\alpha\beta} \big ] = {}& 2i \,\pr_{(\al\dal}
\lambda_{\beta)} \, , \qquad 
\big [ \bQ_{\smash{\dal}}  ,  \DD  \big ] =  \vep^{\be\al}
\pr_{\al \dal} \lambda_{\be} \, , \cr
\big \{ \bQ_{\smash {\dal}}  , \mu_\al \big \} = {}& i \, \vep^{\be\gamma}
\pr_{\gamma \dal} f_{\al\be} + \pr_{\alpha\dal} \DD \, . \cr}
}

For each chiral multiplet there is a corresponding anti-chiral partner obtained
by conjugation when $(\vphi,\psi_\alpha,F)\to ({\bvphi},\bpsi{}_\dal,{\overline F})$,
$(\lambda_\al,f_{\alpha\beta},\DD,\mu_\alpha) \to (\blam_\dal,
{\overline f}{}_{\smash{\dal\dbe}},{\overline \DD},\bmu{}_{\smash{\dal}})$ and when the
$R$-charges change sign.

For the spinor multiplet, with transformations given by \QVfree, \QbarV\ and their
conjugates, we may impose the reality condition
\eqn\realD{
\DD = {\overline \DD}\, .
}
By considering $[Q_\alpha,\DD]$ we must then have
\eqn\mul{
\mu_\alpha = i \, \vep^{\dbe\dal}\pr_{\al\dal}\blam_{\smash{\dbe}} \, ,
}
and using this to calculate $\{\mu_\al,\bQ_{\smash{\dal}}\}$ and
$\{\mu_\al,S^\beta\} = 2 \delta_\al{\!}^\be \vep^{\dbe\dal}\{\bQ_{\smash{\dal}},
\blam_{\smash{\dbe}}\}$ it is also necessary for consistency that
\eqn\consis{
\vep^{\beta\gamma}\pr_{\smash{\gamma\dal}} f_{\alpha\beta} +
\vep^{\dga\dbe} \pr_{\smash{\alpha\dga}} {\overline f}{}_{\smash{\dal\dbe}} = 0 \, , \qquad
r=1 \, .
}
The equation for $f_{\alpha\beta},{\overline f}{}_{\smash{\dal\dbe}}$ is identical
with the abelian Bianchi identity for a field strength
$F_{\smash{\al\dal,\beta\dbe}}= \vep_{\alpha\beta}\,{\overline f}{}_{\smash{\dal\dbe}}
+ \vep_{\smash{\dal\dbe}}\, f_{\alpha\beta}$ and the condition $r=1$, ensuring
$f_{\alpha\beta},{\overline f}{}_{\smash{\dal\dbe}}$ and $\DD$ have vanishing $R$-charge, 
shows that no anomalous dimensions are possible with this restriction. 
The requirement \realD\ of course ensures that 
the chiral spinor multiplet and its anti-chiral conjugate form the superconformal 
multiplet for a gauge field.

For the free chiral scalar field  we have
\eqn\free{
F = 0  \quad \Rightarrow \quad \vep^{\be\al} \pr_{\al \dal} \psi_{\be} = 0 \, ,
\ \  \pr^2 \vphi = 0 \, , \quad r =  {\ts {2\over 3}} \, ,
}
as a consequence of the algebra, \QSfree, \QbarS. For a free spinor multiplet from
\QVfree, \QbarV
\eqn\Vfree{
\DD = \mu_\alpha = 0 \quad \Rightarrow \quad  \vep^{\be\al}
\pr_{\al \dal} \lambda_{\be} = 0 \, , \quad \vep^{\be\al}
\pr_{\al \dal} f_{\be\gamma} = 0 \, , \quad r=1 \, ,
}
which clearly are in accord with \consis.

For the construction of a superconformal index as described in the introduction
we identify in \QQH
\eqn\freeQQ{
\Q = \bQ_1 \, , \qquad \Q^+ = - {\bar S}{}^1 \, , \qquad \H = H - 2 {\bar J}_3 -
{\ts {3\over 2}} R \, .
}
The commuting operators formed from the generators of the superconformal group
$SU(2,2|1)$ and which form the generators of the subgroup $SU(2,1)$ as in \suto\ are 
then
\eqn\subg{
\P_\alpha = \half P_{\alpha 2} \, , \qquad {\bar \P}^\beta = - \half K^{2\beta} \, ,
}
and since $[\P_\alpha , {\bar \P}^\beta ] = M_\alpha{}^{\! \beta} + 
\de_\alpha{}^{\! \beta} ( H + {\bar J}_3 ) $ we have
\eqn\resR{
\R = R + 2 {\bar J}_3 + {\ts{2\over 3}} \H \, ,
}
as in \RR.
 
For free fields it is then straightforward to find the results for the index as defined
in \indt. For the chiral scalar and its conjugate then $[\Q, \Q^+, \vphi]=0, \
\{ \Q, \Q^+, \bpsi_2 \} = 0$ so that the subspace annihilated by $\Q, \Q^+$, and
belonging to the kernel of $\H$, has a basis
\eqn\cfree{
\V_S = \big \{ P_{12}{\!}^n P_{22}{\!}^m |\vphi\rangle, 
P_{12}{\!}^n P_{22}{\!}^m )| \bpsi_2\rangle \big \}  \, , \quad
n,m=0,1,2, \dots \, .
}
where $\R$ has eigenvalues $({2\over 3}+n+m, {4\over 3}+n+m)$ and $2J_3$
$(n-m,n-m)$. Hence evaluating \indt\ on the space spanned by $\V_S$ gives
\eqn\indc{
{\rm str}_{\V_S} \big ( t^{\R} x^{2J_3} \big )
=  {t^{2\over 3}-  t^{4\over 3} \over ( 1 - t x )(1-t x^{-1})} \, ,
}
where the two terms arise from the chiral and anti-chiral fields respectively. 
For the free vector multiplet $\{\Q, \Q^+, \lambda_\alpha \} = 0, \
[ \Q, \Q^+, {\overline f}_{22} ] = 0$ but taking into account the 
equation of motion 
\eqn\dir{
\pr_{22} \lambda_1 = \pr_{12}  \lambda_2 \, .
}
the corresponding basis has the form
\eqn\vfree{
\V_V = \big \{ P_{12}{\!}^n P_{22}{\!}^m |\lambda_1\rangle,
P_{22}{\!}^m |\lambda_2\rangle,
P_{12}{\!}^n P_{22}{\!}^m )| {\overline f}_{22}\rangle \big \}  \, , \quad
n,m=0,1,2, \dots \, .
}
Hence
\eqn\indv{\eqalign{
{\rm str}_{\V_V} \big ( t^{\R} x^{2J_3} \big )
={}& - {t \, x  \over (1 - t x)(1-t x{-1}) }
-  {t \, x^{-1} \over 1 - t x^{-1} }  + {t^2  \over (1 - t x)(1-t x^{-1}) }\cr
={}& {2t^2 - t \, \chi_{2}(x)  \over ( 1 - t x )(1-t x^{-1})} \, , \qquad
\chi_{2}(x) = x + x^{-1} \, . }
}

These results correspond to appropriate $SU(2,1)$ characters as shown in appendix A.
If the chiral field $\vphi$ forms a representation space for a representation $R_S$ 
of a internal symmetry group $\G$ while its anti-chiral partner belongs to the conjugate 
representation ${\bar R}_S$, and the  vector multiplet transforms under the 
self-conjugate  representation $R_V$,then \indc\ and \indv\ can be extended to
\eqn\indexinter{\eqalign{
i_S(p,q,g)={}&{1\over (1-p)(1-q)}\big((p\,q)^{{1\over 3}}\, 
\chi_{\G,R_S}^{\vphantom g}(g)
-(p\, q)^{{2\over 3}}\, \chi_{\G,{\bar R}_S}^{\vphantom g}(g)\big)\, ,\cr
i_V(p,q,g)={}&-\bigg({p\over 1-p}+{q\over 1-q}\bigg)
\chi_{\G,R_V}^{\vphantom g}(g)\, ,}
}
where we introduce the variables $p=tx, q=tx^{-1}$ as in \imat\ and 
$\chi_{\G,R_S}^{\vphantom g}(g)$, $\chi_{\G,{\bar R}_S}^{\vphantom g}(g)$ and
$\chi_{\G,R_V}^{\vphantom g}(g)$ are
corresponding group characters evaluated at $g\in \G$. The general expression in \ione\
is an extension to take into account general $R$-charges for chiral fields.

\newsec{Indices for Seiberg and Kutasov-Schwimmer Duality}

For application to Seiberg duality \Sei, we first consider the usual
$\N=1$ SQCD electric theory with the overall
symmetry group $\G_E=U(1)_R\times U(1)_B\times SU(N_f)\times SU(N_f)\times
SU(N_c)$, where the generator of $U(1)_B$ is the baryon number charge and $U(1)_R$
is generated by the $R$-charge and is part of the superconformal group at a fixed
point, $SU(N_f)\times SU(N_f)$
is the flavour symmetry group while $SU(N_c)$ is the colour gauge group.
For such supersymmetric versions of QCD there are two chiral scalar
multiplets $Q$, ${\tilde Q}$, belonging the fundamental $f$, 
anti-fundamental $\bar f$ representations of $SU(N_c)$, each carrying baryon
number,  and a vector multiplet $V$, in the adjoint. 
The representation content for all fields is detailed in Table 1,
where we have defined
\eqn\defN{
{\tilde N}_c =N_f-N_c\, .
}

\medskip

\vbox{
\hskip1.5cm Table 1: Seiberg Electric Theory
\nobreak

\hskip1.5cm
\vbox{\tabskip=0pt \offinterlineskip
\hrule
\halign{&\vrule# &\strut \ \hfil#\  \cr
height2pt&\omit&&\omit&&\omit&&\omit&& \omit && \omit &\cr
&\   Field  \ \hfil   && \    $SU(N_c)$  \  && \  $SU(N_f)$ \
&& \ $SU(N_f)$ \  && \ $U(1)_B$ \ &&  \ $U(1)_R$ \  &\cr
height2pt&\omit&&\omit&&\omit&&\omit&& \omit && \omit &\cr
\noalign{\hrule}
height2pt&\omit&&\omit&&\omit&&\omit&& \omit && \omit & \cr
& \ $Q$  \ \hfil &&  $f$  \hfil     &&    $f$     \hfil            && 
$1$ \hfil
 &&  $1$    \hfil    &&  ${\tilde N}_c/N_f$   \hfil        & \cr
& \ ${\tilde Q}$  \ \hfil &&  ${\bar f}$  \hfil     &&    $1$     \hfil 
&&  ${\bar f}$ \hfil
&&  $-1$    \hfil    &&  ${\tilde N}_c/N_f$   \hfil        & \cr
& \ $V$  \ \hfil &&  ${\rm adj.}$  \hfil     &&    $1$     \hfil          
 &&  $1$ \hfil
&&  $0$    \hfil    &&  $1$   \hfil        & \cr
height2pt&\omit&&\omit&&\omit&&\omit&& \omit &&\omit &\cr
}
\hrule}
}

The characters $\chi_R(g)$ for $g\in SU(N_c)$ and $\chi_R(h)$ for 
$h\in SU(N_f)\times  SU(N_f)$ are functions of
the complex eigenvalues of $g,h$ for which we adopt the abbreviated notation,
\eqn\defyz{
\z = (z_1,\dots , z_{N_c}) \, , \ {\ts \prod_i} z_i = 1\, , \quad
\y = (y_1,\dots , y_{N_f}) \, , \  {\tilde \y} = ({\tilde y}_1,\dots , {\tilde y}_{N_f}) \, ,
{\ts \prod_i} y_i = {\ts \prod_i} {\tilde y}_i  = 1\, .
}
For $SU(n)$ the required characters, as functions of $\x=(x_1,\dots,x_n)$ with $\prod_i x_i =1$,
are then
\eqn\expforch{\eqalign{
\chi_{SU(n),f}(\z)={}& p_{n}(\x) \equiv {\ts \sum_{j=1}^n}\, x_i\,, \qquad
\chi_{SU(n),{\bar f}}(\x) =  p_{n}(\x^{-1} ) \, , \cr
\chi_{SU(n),\rm adj.}(\x)={}& \sum_{1\leq i,j\leq n} x_i/x_j-1
= p_{n} (\x) p_{n}(\x^{-1} )-1\, ,}
}
using the notation $ \x^{-1} = (x_1{\!}^{-1},\dots , x_{n}{\!}^{-1} ) $.

Applying \expforch\ for $SU(N_c)$ and $SU(N_f)$ the expression given by \ione\
for the single particle index, with $v$ corresponding to $U(1)_B$, becomes
\eqn\seiel{
\eqalign{
i_E & (p,q,v,\y,{\tilde \y},\z)\cr
={}& -\bigg({p\over 1-p}+{q\over 1-q}\bigg)\Big (p_{N_c}(\z)\, p_{N_c}(\z^{-1})-1\Big )\cr
&{} +{1\over (1-p)(1-q)}\bigg((p\,q)^{{1\over 2}r} \, v \, p_{N_f}(\y)\, p_{N_c}(\z)
- (p\,q)^{1- {1\over 2}r} \, {1 \over v}\, p_{{N_f}}(\y^{-1})\, p_{{N_c}}(\z^{-1})\cr
\noalign{\vskip -4pt}
&\qquad\qquad\qquad\qquad
+ (p\,q)^{{1\over 2}r}\, {1\over v}\, p_{{N_f}}({\tilde \y}\,) \, p_{{N_c}}(\z^{-1})
- (p\,q)^{1- {1\over 2}r} \, v\, p_{N_f}({\tilde \y}^{-1})\, p_{N_c}(\z)\bigg) \, ,}
}
where
\eqn\defal{
r = 1 - {N_c\over  N_f}\,.
}

For the dual magnetic theory, whereby the overall symmetry group becomes
$\G_M=U(1)_R\times U(1)_B\times SU(N_f)\times SU(N_f)\times
SU({\tilde N}_c)$ with ${\tilde N}_c$ as in \defN, we have, 
again, two fundamental scalar multiplets $q$, ${\tilde q}$, a
$SU({\tilde N}_c)$ adjoint vector multiplet ${\tilde V}$
along with an extra colour singlet scalar multiplet $M$ with representations 
and $R$-charges as in Table 2. The consistency of the choices in Tables 1
and 2 is determined by applying 't Hooft anomaly  matching conditions.

\bigskip

\vbox{
\hskip1.5cm Table 2: Seiberg Magnetic Theory
\nobreak

\hskip1.5cm
\vbox{\tabskip=0pt \offinterlineskip
\hrule
\halign{&\vrule# &\strut \ \hfil#\  \cr
height2pt&\omit&&\omit&&\omit&&\omit&& \omit && \omit &\cr
&\   Field  \ \hfil   && \    $SU({\tilde N}_c)$  \  && \  $SU(N_f)$ \
&& \ $SU(N_f)$ \  && \ $U(1)_B$ \ &&  \ $U(1)_R$ \  &\cr
height2pt&\omit&&\omit&&\omit&&\omit&& \omit && \omit &\cr
\noalign{\hrule}
height2pt&\omit&&\omit&&\omit&&\omit&& \omit && \omit & \cr
& \ $q$  \ \hfil &&  $f$  \hfil     &&    ${\bar f}$     \hfil          
 &&  $1$ \hfil
 &&  $N_c/{\tilde N}_c$    \hfil    &&  $N_c/N_f$   \hfil        & \cr
& \ ${\tilde q}$  \ \hfil &&  ${\bar f}$  \hfil     &&    $1$     \hfil 
&&  $f$ \hfil
&&  $-N_c/{\tilde N}_c$    \hfil    &&  $N_c/N_f$   \hfil        & \cr
& \ $\tilde V$  \ \hfil &&  ${\rm adj.}$  \hfil     &&    $1$     \hfil         
  &&  $1$ \hfil
&&  $0$    \hfil    &&  $1$   \hfil        & \cr
& \ $M$  \ \hfil &&  $1$  \hfil     &&    $f$     \hfil            && 
${\bar f}$ \hfil
&&  $0$    \hfil    &&  $2{\tilde N}_c/N_f$   \hfil        & \cr
height2pt&\omit&&\omit&&\omit&&\omit&& \omit &&\omit &\cr
}
\hrule}
}

Applying \ione\ the single particle index for the magnetic theory becomes,
in a similar fashion to \seiel, but, for characters for $SU(\wtN_c)$, replacing
$\z$ by ${\tilde \z}$
\eqnn\seim
$$\eqalignno{
i_M& (p,q,v,\y,{\tilde \y},{\tilde \z}) \cr
={}& -\bigg({p\over 1-p}+{q\over 1-q}\bigg)
\Big (p_{{\tilde N}_c}({\tilde \z})\, p_{{\tilde N}_c}({\tilde \z}^{-1})-1\Big )\cr
&+{1\over (1-p)(1-q)}\bigg((p\,q)^{{1\over 2}(1-r)} \, {\tilde v} \, p_{N_f}(\y)\, 
p_{{\tilde N}_c}({\tilde \z})
- (p\,q)^{{1\over 2}(1+r)} \, {1 \over {\tilde v}}\, p_{{N_f}}(\y^{-1})\, 
p_{{\tilde N}_c}({\tilde \z}^{-1})\cr
\noalign{\vskip -4pt}
&\qquad\qquad\qquad\qquad
+ (p\,q)^{{1\over 2}(1-r)}\, {1\over {\tilde v}}\, p_{{N_f}}({\tilde \y}\,) \, 
p_{{\tilde N}_c}({\tilde \z}^{-1})
- (p\,q)^{{1\over 2}(1+r)} \, {\tilde v}\, p_{N_f}({\tilde \y}^{-1})\, 
p_{{\tilde N}_c}({\tilde \z}) \cr
\noalign{\vskip -4pt}
&\qquad\qquad\qquad\qquad
+ (p\,q)^r\, p_{{N_f}}({\y})\, p_{{N_f}}({\tilde \y}^{-1})
- (p\,q)^{1-r}\, p_{N_f}({\y}^{-1})\, p_{N_f}({\tilde \y})\bigg) \, , & \seim }
$$
with $r$ as in \defal\ and the $U(1)_B$ assignments requiring 
\eqn\vv{
{\tilde v}^{{\tilde N}_c}  = v^{N_c} \, .
}

For Kutasov-Schwimmer dual models \KS, the overall symmetry groups are similar to
the Seiberg dual theories considered above but there are additional chiral matter 
fields. In the electric theory there is an extra scalar multiplet $X$ transforming 
according to the adjoint for $SU(N_c)$.
For the dual magnetic theory the $SU({\tilde N}_c)$ gauge group now has
\eqn\defNk{
{\tilde N}_c = kN_f-N_c \, ,  \quad \hbox{for} \quad k=1,2,\dots \, ,
}
and there is also an extra $SU({\tilde N}_c)$ adjoint scalar multiplet $\tilde X$ 
along with now $k$ gauge singlet scalar multiplets, $M_j$, $j=1,\dots,k$.
For $k=1$, these examples reduce to the Seiberg dual theories as 
$X, \,{\tilde X}$ then decouple. The field content in the electric and 
magnetic theories are outlined in Tables 3 and 4.

\medskip

\vbox{
\hskip1cm Table 3: Kutasov-Schwimmer Electric Theory
\nobreak

\hskip1cm
\vbox{\tabskip=0pt \offinterlineskip
\hrule
\halign{&\vrule# &\strut \ \hfil#\  \cr
height2pt&\omit&&\omit&&\omit&&\omit&& \omit && \omit &\cr
&\   Field  \ \hfil   && \    $SU(N_c)$  \  && \  $SU(N_f)$ \
&& \ $SU(N_f)$ \  && \ $U(1)_B$ \ &&  \ $U(1)_R$ \  &\cr
height2pt&\omit&&\omit&&\omit&&\omit&& \omit && \omit &\cr
\noalign{\hrule}
height2pt&\omit&&\omit&&\omit&&\omit&& \omit && \omit & \cr
& \ $Q$  \ \hfil &&  $f$  \hfil     &&    $f$     \hfil            && 
$1$ \hfil
 &&  $1$    \hfil    &&  $\ts 1-{2\over k+1}{N_c\over N_f}$   \hfil       
& \cr
& \ ${\tilde Q}$  \ \hfil &&  ${\bar f}$  \hfil     &&    $1$     \hfil 
&&  ${\bar f}$ \hfil
&&  $-1$    \hfil    &&  $\ts 1-{2\over k+1}{N_c\over N_f}$   \hfil       
& \cr
& \ $V$  \ \hfil &&  ${\rm adj.}$  \hfil     &&    $1$     \hfil          
 &&  $1$ \hfil
&&  $0$    \hfil    &&  $1$   \hfil        & \cr
& \ $X$  \ \hfil &&  ${\rm adj.}$  \hfil     &&    $1$     \hfil          
 &&  $1$ \hfil
&&  $0$    \hfil    &&  $\ts {2\over k+1}$   \hfil        & \cr
height2pt&\omit&&\omit&&\omit&&\omit&& \omit &&\omit &\cr
}
\hrule}
}

This time the electric theory single particle index is given by,
\eqn\ksel{
\eqalign{
i_E & (p,q,v,\y,{\tilde \y},\z)\cr
={}& -\bigg({p\over 1-p}+{q\over 1-q} -{1 \over (1-p)(1-q)}
\big((p\,q)^{s}- (p\,q)^{1-s}\big)
\bigg)\Big (p_{N_c}(\z)\, p_{N_c}(\z^{-1})-1\Big )\cr
&{} +{1\over (1-p)(1-q)}\bigg((p\,q)^{{1\over 2}r} \, v \, p_{N_f}(\y)\, p_{N_c}(\z)
- (p\,q)^{1- {1\over 2}r} \, {1 \over v}\, p_{{N_f}}(\y^{-1})\, p_{{N_c}}(\z^{-1})\cr
\noalign{\vskip -4pt}
&\qquad\qquad\qquad\qquad
+ (p\,q)^{{1\over 2}r}\, {1\over v}\, p_{{N_f}}({\tilde \y}\,) \, p_{{N_c}}(\z^{-1})
- (p\,q)^{1- {1\over 2}r} \, v\, p_{N_f}({\tilde \y}^{-1})\, p_{N_c}(\z)\bigg) \, , }
}
where now
\eqn\defbega{
r =1-{2\over k+1}{N_c\over N_f} \, , \qquad s = {1 \over k+1} \, .
}

\medskip

\vbox{
\hskip1mm
Table 4: Kutasov-Schwimmer Magnetic Theory
\nobreak

\hskip -14mm
\vbox{\tabskip=0pt \offinterlineskip
\hrule
\halign{&\vrule# &\strut \ \hfil#\  \cr
height2pt&\omit&&\omit&&\omit&&\omit&& \omit && \omit &\cr
&\   Field  \ \hfil   && \    $SU({\tilde N}_c)$  \  && \  $SU(N_f)$ \
&& \ $SU(N_f)$ \  && \ $U(1)_B$ \ &&  \ $U(1)_R$ \ \hfil &\cr
height2pt&\omit&&\omit&&\omit&&\omit&& \omit && \omit &\cr
\noalign{\hrule}
height2pt&\omit&&\omit&&\omit&&\omit&& \omit && \omit & \cr
& \ $q$  \ \hfil &&  $f$  \hfil     &&    ${\bar f}$     \hfil        
   &&  $1$ \hfil
 &&  $N_c/{\tilde N}_c$    \hfil    &&  $\ts 1-{2\over k+1}{{\tilde N}_c\over N_f}$   \hfil 
      & \cr
& \ ${\tilde q}$  \ \hfil &&  ${\bar f}$  \hfil     &&    $1$     \hfil 
&&  $f$ \hfil
&&  $-N_c/{\tilde N}_c$    \hfil    &&  $\ts 1-{2\over k+1}{{\tilde N}_c\over N_f}$   \hfil 
      & \cr
& \ $\tilde V$  \ \hfil &&  ${\rm adj.}$  \hfil     &&    $1$     \hfil         
  &&  $1$ \hfil
&&  $0$    \hfil    &&  $1$   \hfil        & \cr
& \ $M_j, \ j=1,\dots k$  \ \hfil &&  $1$  \hfil     &&    $f$     \hfil            && 
${\bar f}$ \hfil
&&  $0$    \hfil    &&  $\ts 2-{4\over k+1}{N_c\over N_f}+{2\over k+1}(j-1) $
 \hfil        & \cr
& \ $\tilde X$  \ \hfil &&  ${\rm adj.}$  \hfil     &&    $1$     \hfil         
  &&  $1$ \hfil
&&  $0$    \hfil    &&  $\ts {2\over k+1}$   \hfil        & \cr
height2pt&\omit&&\omit&&\omit&&\omit&& \omit &&\omit &\cr
}
\hrule}
}

The magnetic theory single particle
index involves a sum over contributions corresponding to $M_j$ of
the form $\sum_{j=1}^k (p\,q)^{r+s(j-1)}$ which is easily summed giving 
\eqn\ksm{\eqalign{
\! i_M& (p,q,v,\y,{\tilde \y},{\tilde \z}) \cr
={}& -\bigg({p\over 1-p}+{q\over 1-q}  -{1 \over (1-p)(1-q)}
\big((p\,q)^{s}- (p\,q)^{1-s}\big) \bigg)
\Big (p_{{\tilde N}_c}({\tilde \z})\, p_{{\tilde N}_c}({\tilde \z}^{-1})-1\Big )\cr
&+{1\over (1-p)(1-q)}\bigg((p\,q)^{s-{1\over 2}r} \, {\tilde v} \, p_{N_f}(\y)\, 
p_{{\tilde N}_c}({\tilde \z})
- (p\,q)^{1-s +{1\over 2}r} \, {1 \over {\tilde v}}\, p_{{N_f}}(\y^{-1})\, 
p_{{\tilde N}_c}({\tilde \z}^{-1})\cr
\noalign{\vskip -4pt}
&\qquad\qquad\qquad\quad
+ (p\,q)^{s-{1\over 2}r}\, {1\over {\tilde v}}\, p_{{N_f}}({\tilde \y}\,) \, 
p_{{\tilde N}_c}({\tilde \z}^{-1})
- (p\,q)^{1-s+{1\over 2}r} \, {\tilde v}\, p_{N_f}({\tilde \y}^{-1})\, 
p_{{\tilde N}_c}({\tilde \z})  \cr
\noalign{\vskip -1pt}
&\qquad\qquad\qquad\quad
+ {1- (p\,q)^{1-s} \over 1- (p\,q)^s}
\Big( (p\,q )^r \, p_{{N_f}}({\y})\, p_{{N_f}}({\tilde \y}^{-1})
- (p\,q )^{2s-r}  \, p_{N_f}(\y^{-1})\, p_{N_f}({\tilde \y})\Big)\bigg) \, .}
}
with the definitions \defbega\ and requiring \vv\ once more. When $k=1$,
$s=\half$ and \ksel\ and \ksm\ reduce to \seiel\ and \seim.

There are important differences between the Seiberg dual theories and those
described by Kutasov and Schwimmer, except in the special case $k=1$ when
the operators $X,{\tilde X}$ decouple. In the former case there is no superpotential
and so no operator relations to take into account. Requiring the colour singlet 
operators $Q{\tilde Q}$ and $q{\tilde q}$ to both satisfy the superconformal 
unitarity bound requires in \seiel\ and \seim\ that  $r, 1-r \ge {1\over 3}$ which 
corresponds, using \defal, to the conformal window ${3\over 2}N_c \le N_f \le 3N_c$.
In  the Kutasov-Schwimmer electric theory  the corresponding condition for the 
operator $Q{\tilde Q}$ also gives $r \ge {1\over 3}$ or with \defbega\  
$N_f \ge 3N_c/(k+1)$. In the magnetic theory there is no similar restriction for 
$q{\tilde q}$ since the superpotential implies that it satisfies
operator relations in this case.

\newsec{Large $N_f$, $N_c$ Limits}

We now show that  the multi-particle index  given by \itwo\ 
with $i$ replaced by $i_E (p,q,v,\y,{\tilde \y},\z)$ with $G=SU(N_c)$ 
and also by  $i_M (p,q,v,\y,{\tilde \y},\z)$ with $G=SU({\tilde N}_c)$ 
agree  in the large $N_c$ limit, requiring $N_f/N_c$ fixed so that
${\tilde N}_c$ is also large. This holds in the general Kutasov-Schwimmer
dual theories which includes the Seiberg dualities as a special case.

Each single particle index above, \seiel, \seim, \ksel\ and \ksm,
may be expressed in the generic form,
\eqn\genfor{
i(\t,\z)=f(\t)\big (p_N(\z)p_{N}(\z^{-1})-1 \big )+g(\t) p_N(\z)+
{\bar g}(\t)p_{N}(\z^{-1}) +h(\t)\, ,
}
for  $f,g,{\bar g},h$ functions of appropriate variables $\t$ and 
$\z=(z_1,\dots,z_N)$. Inserting $i(\t,\z)$ into \itwo, with $G=SU(N)$, the leading 
term in the large $N$ expansion may be obtained by extending the methods used in \FN\
for $g={\bar g}=0$. An alternative approach following \mald\ is also discussed
subsequently.

The method in \FN\ relies on the critical observation that
power symmetric polynomials are orthogonal up to contributions which disappear
in the large $N$ limit.
For $p_N(\z^n)=\sum_{i=1}^N z_i{}^n$,
power symmetric polynomials, which are labelled  by 
$\ua =(a_1,{a_2},\dots)$ $a_i = 0,1,\dots$, are defined by
\eqn\powersym{
p_\ua(\z)=p_{({a_1},{a_2},\dots)}(\z)= p_N(\z)^{a_1}p_N(\z^2)^{a_2}\cdots
\, .
}
These obey the orthogonality relation,
\eqn\orth{
\int_{SU(N)} \!\!\!\!\! \d \mu(\z)\, p_\ua(\z)\,p_{\ua'}(\z^{-1})
=z_\ua  \, \de_{\ua \ua'} \, , \quad |\ua| , |\ua'|<N \, ,
}
where
\eqn\defztau{
z_\ua =z_{(a_1,{a_2},\dots)}=\prod_{n\geq 1}n^{a_n} a_n! \, , \qquad
|\ua | = {\ts \sum_n} n a_n \, .
}
In consequence \orth\ becomes exact for any $\ua,\ua'$ in the large $N$ limit.

This result may now be used to evaluate
\eqn\ithree{
\I(t) = \int_{SU(N)} \!\!\!\!\! \d \mu(\z)\, \exp\bigg(\sum_{n=1}^\infty{1\over n}
i(\t^n,\z^n)\bigg) \, ,
}
by expanding the exponential
\eqn\plethe{\eqalign{
&\exp\bigg(\sum_{n=1}^\infty{1\over n}i(\t^n,\z^n)\bigg)\cr
&=
\exp\bigg(\sum_{n=1}^\infty{1\over n}\big({-f(\t^n)}+h(\t^n)\big)\bigg)
\prod_{n=1}^\infty
\sum_{a_n,b_n,{\bar b}_n\geq 0}{1\over n^{a_n+b_n+{\bar b}_n}\, a_n! \,b_n! \,
{\bar b}_n!}     \cr
&\qquad \qquad\qquad \qquad \qquad
\times f(\t^n)^{a_n}\,g(\t^n)^{b_n}\, {\bar g}({\t}^n)^{{\bar b}_n}\,
p_N(\z^n)^{a_n+b_n}\, p_N(\z^{-n})^{a_n+{\bar b}_n} \, ,}
}
so that, applying \orth, \defztau\ in \ithree, the $SU(N)$ integral gives
\eqn\indexinfNn{\eqalign{
\I(\t) \simeq {}&\exp\bigg(
\sum_{n=1}^\infty{1\over n}\big({-f(\t^n)}+h(\t^n)\big)\bigg)\cr
&\times\prod_{n=1}^\infty \sum_{a_n,b_n\geq 0}
{(a_n+b_n)!\over n^{b_n}\, a_n!\,  b_n!{}^2} \,  f(\t^n)^{a_n}
\big( g(\t^n)\,{\bar g}({\t}^n)\big){}^{b_n} \, ,}
}
where the right hand side is exact so long as \orth\ holds and so \indexinfNn\ is 
valid, up to contributions which are negligible for large $N$. 
Using $\sum_{r=0}^\infty \left({r+s\atop r}\right)x^r=1/(1-x)^{s+1}$ and
$\sum_{s=0}^\infty {1\over n^s s!} \, {y^s \over (1-x)^{s+1}} = {1\over 1-x}\, 
\exp( {1\over n}\, {y\over 1-x})$ we then easily obtain
\eqn\indexinfN{
\I(\t) \simeq \exp\bigg(
\sum_{n=1}^\infty{1\over n}
\Big(    {g(\t^n)\,{\bar g}({\t}^n)\over 1-f(\t^n)} - f(\t^n)+h(\t^n)
\Big)\bigg)\prod_{n=1}^{\infty}{1\over 1-f(\t^n)}\, .}
This result gives the leading large $N$ form for $\I(t)$ if we assume
$g, {\bar g}$ are both ${\rm O}(N)$ and $h$ is ${\rm O}(N^2)$.

Alternatively we can also show how conventional large $N$ techniques give the
same result \indexinfN. For $z_i = e^{i\theta_i}$ the invariant integration over
$SU(N)$ has the form
\eqn\intth{
\int_{SU(N)} \!\!\!\!\! \d \mu(\z)\, = {1\over N} \, {1\over (2\pi)^{N-1}} 
\int_{\lower2pt\hbox{$\scriptstyle-\pi\le \theta_1\le \theta_2\le \dots \le 
\theta_{N-1}\le \pi$}} 
{\hskip -2cm {\ts{\prod_{i=1}^{N-1}}} \d \theta_i} \qquad
\prod_{i<j} 4 \sin^2 \half ( \theta_i - \theta_j) \,,
}
where we impose $\sum_i \theta_i = 0$. The basic integral \ithree\ then becomes
\eqn\ifour{
\I(\t) = {1\over N}\, {1\over (2\pi)^{N-1}} 
\int_{\lower2pt\hbox{$\scriptstyle-\pi\le \theta_1\le \dots \le
\theta_{N-1}\le \pi$}}
{\hskip -2cm {\ts{\prod_{i=1}^{N-1}}} \d \theta_i} \qquad
e^{-S(\t,\theta)} \, ,
}
for
\eqn\act{ 
S(\t,\theta) =  \sum_{n=1}^\infty {1\over n} \bigg \{
\big ( 1 - f(\t^n) \big ) \sum_{i \ne j} e^{in (\theta_i - \theta_j)} 
- g(\t^n) \sum_i e^{-in \theta_i} - {\bar g}(\t^n) \sum_i e^{in \theta_i}
- {\hat h}(\t^n) \bigg \} \, ,
}
defining for convenience ${\hat h}= h-f$.
In the large $N$ limit we assume $\theta_i \to \theta(i/N)$, a continuous monotonic
function such that $ \sum_i f(\theta_i) \to N \int_0^1 \d x f(\theta(x))$. 
In \ifour\ the product of $\d\theta_i$
integrals then becomes a functional integral $\d[\theta]$. The asymptotic
evaluation is obtained by introducing instead of $\theta(x)$ 
a density function $\rho(\theta)$ defined in terms of $\theta(x)$ by
\eqn\defrho{
{\d x \over \d \theta} = \rho(\theta) \, ,
}
and then defining
\eqn\mom{
\rho_n = N \int_{-\pi}^\pi \!\!\! \d \theta \, \rho(\theta) \, e^{in \theta} \, , 
\qquad \rho_0 = N \, ,
}
we assume
\eqn\intp{
\int_{\theta'>0} \!\!\!\!\! \d[\theta] \to \int \d[\rho] = 
\int \prod_{n\ge 1} {n\over \pi} \d^2 \rho_n \,,
}
normalising to unit group volume. Letting
\eqn\Srho{
S(\t,\theta) \to {\tilde S}(\t,\rho)
= \sum_{n=1}^\infty {1\over n} \Big \{
\big ( 1 - f(\t^n) \big )\, \rho_n \rho_{-n}
- g(\t^n)\, \rho_{-n} - {\bar g}(\t^n)\, \rho_n - h(\t^n) \Big \} \, ,
}
we obtain
\eqn\asymI{
\I(\t) \simeq \int \d[\rho] \, e^{-{\tilde S}(\t,\rho)} \, ,
}
which is a straightforward Gaussian functional integral, assuming $1-f(t)>0$.
The saddle points are
\eqn\sad{
{\hat \rho}_n = { g(\t^n) \over 1 - f(\t^n)} \, , \quad
{\hat \rho}_{-n} = { {\bar g}(\t^n) \over 1 - f(\t^n)} \, , \quad n=1,2,\dots \, ,
}
and it is easy to see that \asymI\
reproduces the leading expression shown in \indexinfN, although it is not so 
evident that this result is exact for the first few terms in an expansion.

We now apply \indexinfN\  to verify  that it gives the same expression for 
both dual electric and magnetic theories considered in the previous section.
Since the Seiberg dual theories are a special case of those considered by
Kutasov and Schwimmer we focus on the latter. Comparing \genfor\ with \ksel\
and \ksm\ it is easy to see that $f$ in \genfor\  is the same in both cases and that
\ksel, \ksm\ give
\eqn\omf{
1 - f(p,q) = {(1+ (p\, q)^{1-s} ) ( 1- (p\,q)^s) \over (1-p)(1-q)} \, .
}
We may then read off from \ksel, comparing with \genfor,
\eqn\kfghel{\eqalign{
g_E(p,q,v,\y,{\tilde \y})&{}
={v\over (1-p)(1-q)} \Big ((p\,q)^{{1\over 2}r} \, p_{N_f}(\y) 
- (p\,q)^{1- {1\over 2}r} \, p_{N_f}({\tilde \y}) \Big ) \, ,\cr
{\bar g}_E(p,q,v,\y,{\tilde \y}) &{} ={v^{-1}\over (1-p)(1-q)}
\Big((p\,q)^{{1\over 2}r} \, p_{N_f}({\tilde \y}^{-1} )\
- (p\,q)^{1- {1\over 2}r} \, p_{{N_f}}(\y^{-1}) \Big ) \, , \cr
h_E(p,q,\y,{\tilde \y}) &{}=0\, ,\cr}
}
and, from \ksm,
\eqn\kfghm{\eqalign{
g_M(p,q,v,\y,{\tilde \y})&{}
={{\tilde v}\over (1-p)(1-q)} \Big ((p\,q)^{s-{1\over 2}r} \, p_{N_f}(\y^{-1}) 
- (p\,q)^{1-s+ {1\over 2}r} \, p_{N_f}({\tilde \y}^{-1}) \Big ) \, ,\cr
{\bar g}_M(p,q,v,\y,{\tilde \y}) &{} ={{\tilde v}^{-1}\over (1-p)(1-q)}
\Big((p\,q)^{s-{1\over 2}r} \, p_{N_f}({\tilde \y} )\
- (p\,q)^{1-s+ {1\over 2}r} \, p_{{N_f}}(\y) \Big ) \, , \cr
h_M(p,q,\y,{\tilde \y}) &{} ={1\over (1-p)(1-q)} {1- (p\,q)^{1-s} \over 1- (p\,q)^s} \cr
\noalign{\vskip -2pt}
&\qquad {}\times \Big( (p\,q )^r \, p_{{N_f}}({\y})\, p_{{N_f}}({\tilde \y}^{-1})
- (p\,q )^{2s-r}  \, p_{N_f}(\y^{-1})\, p_{N_f}({\tilde \y})\Big)
\, , \cr}
}
with the same notation as in \defbega\ and also requiring \vv.
{}From \kfghel\ and \kfghm\ we have
\eqn\gdiff{\eqalign{
& g_E(p,q,v,\y,{\tilde \y})\, {\bar g}_E(p,q,v,\y,{\tilde \y})  -
g_M(p,q,v,\y,{\tilde \y})\, {\bar g}_M(p,q,v,\y,{\tilde \y}) \cr
&{}  = {1 - (p\,q)^{2(1-s)} \over (1-p)^2(1-q)^2} \,
\Big( (p\,q )^r \, p_{{N_f}}({\y})\, p_{{N_f}}({\tilde \y}^{-1})
- (p\,q )^{2s-r}  \, p_{N_f}(\y^{-1})\, p_{N_f}({\tilde \y})\Big) \, ,}
}
and hence
\eqn\relI{
{g_E(p,q,v,\y,{\tilde \y})\, {\bar g}_E(p,q,v,\y,{\tilde \y}) \over 1 - f(p,q)}
= {g_M(p,q,v,\y,{\tilde \y})\, {\bar g}_M(p,q,v,\y,{\tilde \y}) \over 1 - f(p,q)}
+ h_M(p,q,\y,{\tilde \y}) \, .
}
Thus \indexinfN\ demonstrates that the large $N$ limit for the index is the same
in both dual and electric theories. In this limit there is no dependence on the 
$U(1)_B$ variable $v$ since there is no contribution from baryon operators and 
this limit is also insensitive to the precise dual gauge groups.

Applying \indexinfN\ in this case then gives for the index
\eqnn\IE
$$\eqalignno{
I(p,q,v,\y,{\tilde y}) \simeq {}&
\exp \bigg ( \sum_{n=1}^\infty {1\over n} \Big ( 
{g_E(p^n,q^n,v^n,\y^n,{\tilde \y}^n)\, {\bar g}_E(p^n,q^n,v^n,\y^n,{\tilde \y}^n) \over 
1 - f(p^n,q^n)} - f(p^n,q^n) \Big ) \bigg ) \cr
\noalign{\vskip -3pt}
&{}\times \prod_{n=1}^\infty {1\over 1 - f(p^n,q^n)} \, . & \IE  \cr }
$$
The first few terms in the expansion involving operators of low scale dimension 
are then
\eqn\seriesexp{\eqalign{
I(t x, t x^{-1}, v, \y,{\tilde \y})
= {}& 1+  t^{2r} p_{N_f}(\y) p_{{ N_f}}({\tilde \y}^{-1}) + t^{4-2r} 
p_{{N_f}}(\y^{-1}) p_{N_f}({\tilde \y})\cr
&{} - t^2 \big ( p_{N_f}(\y) p_{N_f}(\y^{-1})+  p_{N_f}({\tilde \y}) 
p_{N_f}({\tilde \y}^{-1}) \big ) \cr
&{}+ t^{4s} -  \big (t^{1+2s} - t^{3-2s} \big ) \chi_2(x) + \dots \, , }
}
where $\chi_2(x) = x + x^{-1}$ is a $SU(2)$ character corresponding to operators 
with  $j=\half$. In the Seiberg case, when $s=\half$ and $r$ is given by \defal, 
the results shown in \seriesexp\ are in exact accord with the tables in \romel.  
The expansion of \IE\ neglects contributions from
operators with non-zero baryon charge which first arise at ${\rm O}(t^{N_c r})$.
In \seriesexp\ the expansion clearly generates integer coefficients, as required
in \expI, to this limited order. 
Except for  the Seiberg case the expression for the index may be expected
be modified once
constraints on the operator spectrum arising from the superpotential are
incorporated.

\newsec{Index Matching for $\N=1$ Superconformal $SU(2)$
Gauge Theories with Three Flavours and its Seiberg Dual}

For the Seiberg dual theories analytic proofs of the equality of the
index between the electric and magnetic  theories are possible for
general finite $N_c,N_f$. These depend crucially on the detailed choice
of the dual gauge groups and the assignments of $U(1)_B$ charges and
provide  very non-trivial tests of  duality in this case and also of the framework
described here for calculating the index in these theories.

As a simple example in this section we discuss in some detail the example of dual theories 
when  $(N_c,N_f)=(2,3)$. There are various simplifications in this case. Since for 
$N_c=2$ there is no distinction between the fundamental representation and its conjugate 
the flavour symmetry group extends $U(1)_B \times SU(3) \times SU(3) \to SU(6)$.  
In the electric theory $Q^a = (Q^i, {\tilde Q}_i)$
forms the six dimensional fundamental representation while in the magnetic dual theory
$q^{ab} = (\epsilon^{ijk}q_k, \epsilon_{ijk}{\tilde q}^k, M^i{\!}_j, -M^j{\!}_i)$ forms the 
15 dimensional antisymmetric tensor representation $T_A$. The index formulae are then
more simply given in terms of $SU(6)$ characters which depend on
\eqn\defY{
{\rmY} =  (p\,q)^{1\over 6} \big ( v\y, v^{-1} {\tilde \y} \big ) \, , 
\qquad {\ts \prod_{a=1}^6 } u_a = pq \, ,
}
where the rescaling is introduced to ensure $i_E,i_M$ have the form exhibited in \imat, \igauge.
Also in this example ${\tilde N}_c =1$ so the magnetic theory at the superconformal fixed point
is a free theory. From \seiel, since for $N_c=2$ we may take $\z=(z,z^{-1})$, we then have
\eqn\spel{
\eqalign{
i_E (p,q,{\rmY},z)
={}& -\bigg({p\over 1-p}+{q\over 1-q}\bigg)\chi_3(z) \cr
&{} +{1\over (1-p)(1-q)}\big( p_{6} ({\rmY} )
- p\,q\, p_{{6}} ({\rmY}^{-1} ) \big ) \chi_2(z) \, ,  \cr}
}
with the $SU(2)$ characters
\eqn\chit{ 
\chi_3(z) = z^2 +1 + z^{-2} \, , \qquad \chi_2(z) = z + z^{-1} \, .
}
Also from \seim
\eqn\asda{
i_M(p,q,{\rmY})=  {1\over (1-p)(1-q)}\big ( \chi_{SU(6),T_A}^{\vphantom g}
({\rmY}) - p\,q \, \chi_{SU(6),T_A}^{\vphantom g}({\rmY}^{-1} ) \big  )\,,
}
where the character for the antisymmetric tensor representation for $SU(n)$ 
has the form
\eqn\cTA{
\chi_{SU(n),T_A}^{\vphantom g} (\x)  = \sum_{1\leq i<j \leq n} x_i x_j \, , \qquad
\chi_{SU(n),{\bar T}_A}^{\vphantom g}(\x ) = \chi_{SU(n),T_A}^{\vphantom g} (\x^{-1})\, .
}

For $SU(2)$ the invariant measure becomes
\eqn\invtwo{
\int_{SU(2)} \!\!\!\!\! \d \mu(z)\, f(z) =  
-{1\over 4 \pi i}\oint{\d z\over z^3} (1-z^2)^2 f(z) 
= {1\over 2\pi i}\oint{\d z\over z}(1-z^2 )\, f(z) \, ,
}
for any analytic $f(z)=f(z^{-1})$. Hence we may express the index for the electric theory  by
using \igauge\  with \eThe
\eqn\indexe{\eqalign{
{I}_E(p,q,{\rmY})={}&  \int_{SU(2)} \!\!\!\!\! \d \mu(z)\, \exp\bigg(\sum_{n=1}^\infty
{1\over n} i_E \big (p^n,q^n,{\rmY}^n,z^n \big )\bigg) \cr
={}&  - (p;p)\, (q;q) \, {1\over 4 \pi i}\oint{\d z\over z^3} \, 
\theta(z^2;p)\, \theta (z^2;q)\, \I(p,q,{\rmY},z) \, , }
}
where for $|u_a|<1$ the contour may be restricted to the unit circle.
With the aid of \imat\ and \eGam
\eqn\Iin{
\I(p,q,\rmY,z)= \prod_{a=1}^6 \Gamma(u_a z;p,q) \, \Gamma(u_a/z;p,q) \, ,
}
or, since from the definition \eGam,
\eqn\GG{
\Gamma(y;p,q) \, \Gamma(pq/y;p,q) = 1 \, ,
}
then, with the constraint \defY, we may also write \Iin\  in a form involving just 
$\hu=(u_1, \dots, u_5)$ 
\eqn\defl{
\I(p,q,\rmY,z) = \hI(p,q,\hu,z)
= {\prod_{a=1}^5\Gamma(u_a z;p,q)\,\Gamma(u_a/z;p,q)\over 
\Gamma(\lambda z;p,q)\, \Gamma(\lambda/z;p,q)} \,,
\qquad \lambda=\prod_{a=1}^5 u_a\, ,
}
so that, with $|u_a|<1, \ a=1,\dots 5$ and $|\lambda| >pq$,
\eqn\IA{
{I}_E(p,q,{\rmY})= \A(p,q,\hu)  
= - (p;p)\, (q;q) \, {1\over 4 \pi i}\oint{\d z\over z^3} \, 
\theta(z^2;p)\, \theta (z^2;q)\, \hI(p,q,\hu,z) \, .
}

For the magnetic index there is no integration so that \eGam\ gives directly
\eqn\indexm{\eqalign{
{I}_M(p,q,{\rmY})={}&  \exp\bigg(\sum_{n=1}^\infty
{1\over n} i_M \big (p^n,q^n,{\rmY}^n,z^n \big )\bigg) \cr
={}&  \prod_{1\leq a<b \leq 6}\Gamma(u_a u_b;p,q) \cr
={}& {\prod_{1\leq a<b \leq 5}\Gamma(u_a u_b;p,q)\over \prod_{a=1}^5
\Gamma(\lambda/u_a;p,q)} = \B(p,q,\hu)\, , }
}
where in the last line we have used \GG\ again to write the index in terms of $\hu$.

An identity obtained by Spiridonov \spi\ shows that \indexe\ and \indexm\ are 
identical. This result is discussed in appendix E, but here we consider on the
special case for $p=0$, which is known in the relevant literature 
as the Nassrallah-Rahman theorem, and summarise a particular simple proof
which may be generalised to show $\A(p,q,\hu)=\B(p,q,\hu)$.\foot{Even for $p=q=0$,
and taking also $u_6=0$, the identities are not entirely trivial. In this limit
$i_E(0,0,\u,z) = (p_5(\hu)- \lambda ) \chi_2(z)$ and $i_M(0,0,\u)= \sum_{1\le a < b \le 5}
u_a u_b - \sum_{1\le a \le 5} \lambda/u_a$. Hence
$$ \eqalign{
I_E(0,0,\u) ={}& {1\over 2\pi i} \oint {\d z \over z}(1-z^2) \, 
{(1- \lambda z) (1- \lambda/ z) \over \prod_{1\le a \le 5} ( 1-u_a z)(1-u_a/z)}
= \sum_b {(1-\lambda u_b)(1-\lambda/u_b) \over \prod_{a\ne b} (1-u_a u_b)(1-u_a/u_b)}\cr
= {}& {\prod_a(1-\lambda/u_a) \over \prod_{a<b} ( 1-u_a u_b) } = I_M(0,0,\u) \, .}
$$
This result may be expanded in terms of Schur polynomials as
$$
I_E(0,0,\u) = \sum_{n\ge 0} \,  ( s_{(n,n,0,0,0,0)}(\u) + s_{(n-3,n-3,2,2,2,0)}(\u)
- s_{(n-1,n-2,1,1,1,0)}(\u) )   \, ,
$$
where we set $u_6=0$ and the three terms  contribute for $n\ge 0, 5 , 3$ respectively.
This matches the leading terms in the expansion given in \romel.
}

When $p=0$ \indexe\ may be written in the form
\eqn\indexst{
I_E(0,q,\rmY)=(q;q) \, {1\over 4\pi i}\oint{\d z\over z}\, (z^2;q)(z^{-2};q)\, 
\J(q,\hu,z)  = \L (q,\hu) \,,
}
with definitions in \defth, where
\eqn\defIq{
\J(q,\hu,z)={(\lambda z;q)\, (\lambda/z;q)\over \prod_{a=1}^5
(u_a z;q) \, (u_a /z;q)}\, .
}
The Nassrallah-Rahman theorem \Nas\ implies essentially that \indexst\ is equal to
\eqn\anstt{
I_M(0,q,\rmY)={\prod_{a=1}^5(\lambda/u_a;q) \over \prod_{1\leq a<b\leq 5}
(u_au_b;q)}  = \R (q,\hu)  \,.
}
If $u_5=0$ the corresponding integral is a well known result first considered by Askey
and Wilson, see \basic. A simple proof due to Askey for this result was also
extended to the full integral given by \indexst\ and \anstt\ \askey\ and 
involves first finding a $q$-difference relation satisfied by
$\J(q,\hu,z)$, when $u_a \to q u_a$ for a particular $a$ and any $z$ so that it must 
hold for $\L (q,\hu)$ as well. 
The essential requirement is that this is also satisfied by $\R(q,\hu)$. 
The $q$-difference relation is then shown to allow a proof of the identity 
$\L (q,\hu)= \R(q,\hu)$ to be derived from that for some suitable special cases 
for $\hu$.

The required $q$-difference relation is obtained from
\eqn\varcho{
\J(q,qu_1,u_2,\dots,u_5,z)={(1-u_1 z)(1-u_1 /z)\over (1-\lambda z)(1-\lambda/z)} 
\, \J(q,\hu,z)\, ,
}
and then using the identity
\eqnn\fourtid
$$\eqalignno{
& u_2(1- u_1 z)(1- u_1 /z)(1-\lambda u_2 )(1- \lambda/ u_2)
- u_1(1- u_2 z)(1- u_2 /z)(1-\lambda u_1 )(1- \lambda/ u_1) \cr
&= -(u_1 - u_2 )(1- u_1 u_2)(1- \lambda z)(1- \lambda/ z)\,, & \fourtid }
$$
to show that $\J(q,\hu,z)$ satisfies
\eqnn\qdiffo
$$\eqalignno{
& u_2(1-\lambda u_2)(1-\lambda/u_2)\, \J(q,qu_1,u_2,\dots,u_5,z)-
u_1(1-\lambda u_1)(1-\lambda/u_1)\, \J(q,u_1,qu_2,\dots,u_5,z)\cr
&=-(u_1-u_2)(1-u_1u _2)\, \J(q,\hu,z)\, . & \qdiffo }
$$
Clearly from \indexst\  $\L(q,\hu)$ satisfies the same $q$-difference relation.
Also we have from \anstt
\eqn\varcht{
\R(q,qu_1,u_2,\dots,u_5)=\prod_{a=2}^5{1- u_1u_a\over 1-\lambda/u_a}\, \R(q,\hu)\,,
}
and in this case using the identity, for $\lambda$ as in \defl,
\eqn\thrup{\eqalign{
& u_2 ( 1 - \lambda u_2 )\, {\ts \prod_{a \ne 1}} ( 1 - u_1 u_a ) -
u_1 ( 1 - \lambda u_1 )\, {\ts \prod_{a \ne 2}} ( 1 - u_2 u_a ) \cr
& = - (u_1-u_2) ( 1-u_1 u_2)\,  {\ts \prod_{a=3}^5} ( 1 - u_a/\lambda ) \, , \cr}
}
it is then easy to show that, as well as $\L(q,\hu)$,  $\R(q,\hu)$ also satisfies \qdiffo.

For the special case chosen in \askey, $\hu_0=(u,1,-1,q^{1\over 2},-q^{1\over 2})$,
we then have
\eqn\usp{
(z^2;q)(z^{-2};q)\,\J(q,\hu_0,z) = {1\over (1-uz)(1-u/z)} \, , 
}
using the identity $(z;q)(-z;q)(q^{1\over 2}z,q)(-q^{1\over 2}z,q)=(z^2,q)$, 
and it is easy to calculate the contour integral in \indexst\ giving
\eqn\Isp{
\L (q,\hu_0) = {(q,q) \over 2(1-u^2)} \, .
}
The same result holds from \anstt\ for $\R (q,\hu_0)$ using $(-q;q)(q,q^2)=1$.
The $q$-difference relation implies $\L(q,\hu_n)= \R (q,\hu_n)$ for
$\hu_n = (u,q^n,-1,q^{1\over 2},-q^{1\over 2})$. Analyticity ensures that equality
must hold for any $u_1,u_2$ and further similar discussion extends this to
any $\hu$.

\newsec{Index Matching for $\N=1$ Superconformal $SU(N_c)$ Gauge Theories with
$N_f$ Flavours and its Seiberg Dual}

In this section we show how the matching between the
multi-particle indices for the general $(N_c,N_f)$ case of
Seiberg duality boils down to a theorem for the transformations of certain elliptic
hypergeometric integrals, due to Rains \rains.  The exact results here apply
just to the Seiberg dual theories described in section 3.

For the invariant integral over $SU(n)$ of any symmetric function $f(\x)$,
$\x=(x_1, \dots , x_{n})$, we have, equivalent to \intth,
\eqn\invG{
\int_{SU(n)} \!\!\!\! \d \mu(\x)  \, f(\x) = 
{1\over n!}\int_{\Bbb T_{n-1}} \! \prod_{j=1}^{n-1}
{\d x_j\over 2\pi i x_j} \, \Delta(\x)\Delta(\x^{-1}) \, f(\x) 
\bigg |_{\prod_{j=1}^{n} x_j =1 } \, ,
}
for $\Bbb T_{n-1} = S^1 \times \dots \times S^1$ the unit torus and 
where the Vandermonde determinant is, as usual,
\eqn\van{
\Delta(\x)=\prod_{1\leq i< j\leq n}(x_i-x_j)\, .
}

For application here it is convenient to rescale the $SU(N_f) \times SU(N_f)$ variables
\eqn\redefvaro{ 
(p\,q)^{{1\over 2}r}v\,\y\to \y\, , \qquad 
(p\,q)^{-{1\over 2}r} v \, \, {\tilde \y}\to {\tilde \y} \, ,
}
where now 
\eqn\deflamb{
\prod_{j=1}^{N_f} y_j = (p\,q)^{{1\over 2}{\tilde N}_c}v^{N_f} = \lambda^{{\tilde N}_c}  \, , 
\qquad \prod_{j=1}^{N_f} {\tilde y}_j = (p\,q)^{-{1\over 2}{\tilde N}_c}v^{N_f} =
{\tilde  \lambda}{}^{{\tilde N}_c}  \, ,
}
and then \seiel\ becomes
\eqn\ienc{\eqalign{
i_E(p,q,\y,{\tilde \y},\z)={}& -\bigg({p\over 1-p}+{q\over 1-q}\bigg)
\bigg(\sum_{1\leq i,j\leq N_c}z_i/z_j-1\bigg)\cr
&{}+{1\over
(1-p)(1-q)}\sum_{i=1}^{N_f}\sum_{j=1}^{N_c}\Big(\big (y_i-p\,q\,{\tilde y}_i \big )z_j
+ \big ({\tilde y}_i{\!}^{-1}-p\,q\,y_i{\!}^{-1} \big )z_j{\!}^{-1}\Big)\, .
}}
Hence, using \eGam\ and \eThe,
\eqn\psum{\eqalign{
\exp &\bigg(\sum_{n=1}^\infty{1\over n} i_E\big (p^n,q^n,\y^n,{\tilde \y}^n,\z^n\big ) 
\bigg )\cr
= {}& {1 \over  \Delta(\z)\Delta(\z^{-1})}\,  {(p;p)^{N_c-1}\, (q;q)^{N_c-1}\over 
\prod_{1\leq i<j\leq
{N_c}}\Gamma(z_i/z_j,z_j/z_i;p,q)}  \prod_{1\leq i\leq N_f}\! \prod_{1\leq j\leq N_c}
\!\! \Gamma \big (y_i z_j,1/({\tilde y}_i z_j);p,q \big )\, ,}
}
where we use
\eqn\identvan{
\prod_{1\leq i,j\leq N_c\atop i\neq j}(1-z_i/z_j)=\De(\z)\De(\z^{-1})\, ,
}
and adopt the notation
\eqn\mGam{
\Gamma(x_1,\dots,x_n;p,q)=\Gamma(x_1;p,q)\cdots \Gamma(x_n;p,q)\, .
}
Applying \invG\ for $SU(N_c)$ and \psum, the expression \itwo\ for the electric index
becomes
\eqn\indexE{\eqalign{
\!\!\!\!\!\! & I_E(p,q,\y,{\tilde {\y}})_{SU(N_c)}\cr
\!\!\!\!\!\! & = (p;p)^{N_c-1} (q;q)^{N_c-1}
{1\over N_c!} \int \! \prod_{j=1}^{N_c-1} \! {\d z_j\over 2\pi i z_j}
{\prod_{1\leq i\leq N_f}\prod_{1\leq j \leq N_c}
\Gamma \big (y_i z_j,1/({\tilde y}_i z_j);p,q \big )\over \prod_{1\leq i<j\leq N_c}
\Gamma\big (z_i/z_j,z_j/z_i;p,q \big )} \bigg |_{\prod_{j=1}^{N_c}z_j=1}\,,}
}
which is solely in terms of elliptic gamma functions. The denominator in \indexE\
is naturally associated with the root system $A_{N_c-1}$, which is expressible
in terms of orthonormal unit vectors as the $N_c(N_c-1)$ roots 
$\pm(e_i-e_j)$, $1\leq i<j\leq N_c$, where we map the root $e_i-e_j$ to
the $\Gamma$ function depending on $ z_i/z_j$.

For the magnetic dual theory then rewriting \seim\ with the rescaling \redefvaro\
and the definitions \deflamb,
\eqn\im{\eqalign{
&\!\!\!\!\!\!\! i_M(p,q,\y,{\tilde \y},{\tilde \z})\cr
&\!\!\!\!\!\!\!=-\bigg({p\over 1-p}+{q\over 1-q}\bigg)
\bigg(\sum_{1\leq i,j\leq {\tilde N}_c } \!\!\!\!\! \tz_i/\tz_j-1\bigg)\cr
&\!+{1\over (1-p)(1-q)}
\bigg(\sum_{i=1}^{N_f}\sum_{j=1}^{{\tilde N}_c} \Big( \big (\lambda\, {y}_i{\!}^{-1}
-p\,q\,{\tilde \lambda}\, {\tilde y}_i{\! }^{-1}\big )\, \tz_j
+ \big ({\tilde \lambda}^{-1} {\tilde y}_i -  p\,q\,\lambda^{-1} y_i \big ) 
\, \tz_j{\!}^{-1}\Big)\cr
\noalign{\vskip -3pt}
&\qquad\qquad\qquad\qquad
+\sum_{i,j=1}^{N_f}
\Big(y_i\,{\tilde y}_j{\!}^{-1}-p\,q\,y_i{\!}^{-1}\,{\tilde y}_j\Big) \bigg)\,. }
}
Hence following the same route as that leading to \indexE
\eqn\indexM{\eqalign{
I_M & (p,q,\y,{\tilde{\y}})_{SU(\wtN_c)}\cr
 ={}& 
{1\over {\tilde N}_c!} \int \prod_{j=1}^{{\tilde N}_c-1}{\d \tz_j\over 2\pi i \tz_j} \,
\Delta(\z)\Delta(\z^{-1})\exp\bigg(\sum_{n=1}^\infty{1\over n}
i_M(p^n,q^n,\y^n,{\tilde \y}^n,{\tilde \z}^n) \bigg) 
\bigg |_{\prod_{j=1}^{{\tilde N}_c}\tz_j=1}\cr
={}& \prod_{1\leq i,j\leq N_f} \!\! \Gamma(y_i/{\tilde y}_j;p,q)\
(p;p){}^{{\tilde N}_c-1}\, (q;q){}^{{\tilde N}_c-1} \cr
\noalign{\vskip -2pt}
{}&\times {1\over {\tilde N}_c!} \int \prod_{j=1}^{{\tilde N}_c-1}
{\d \tz_j\over 2\pi i \tz_j} \,
{\prod_{1\leq i\leq N_f}\prod_{1\leq j\leq {\tilde N}_c}
\Gamma \big (\lambda \tz_j/y_i,{\tilde \lambda}^{-1}{\tilde y}_i /\tz_j;p,q \big )
\over \prod_{1\leq i<j\leq {\tilde N}_c}\Gamma\big (\tz_i/\tz_j,\tz_j/\tz_i;p,q\big )} 
 \bigg |_{\prod_{j=1}^{{\tilde N}_c}\tz_j=1}  \cr
={}& \prod_{1\leq i,j\leq N_f} \!\! \Gamma(y_i/{\tilde y}_j;p,q)\
I_E (p,q,\lambda \y^{-1} , {\tilde \lambda}{\tilde{\y}^{-1} })_{SU(\wtN_c)}  \,.}
}

The essential requirement is the electric and magnetic theories are identical
at the IR superconformal fixed point so that
\eqn\II{
I_E(p,q,\y,{\tilde {\y}})_{SU(N_c)} = I_M(p,q,\y,{\tilde {\y}})_{SU(\wtN_c)}  \, .
}
The integrals appearing in \indexE\ and \indexM\ are just those
considered by Rains \rains. The right hand side of \indexE\ for $n=N_c-1, \
m= {\tilde N}_c-1$ defines the elliptic hypergeometric integral
$I_{A_n}^{(m)}\big (\y;{\tilde \y}^{-1};p,q\big )$, depending on $(m+n+2)$-dimensional vectors
$\y,{\tilde \y}$. Theorem 4.1 of \rains\ requires
\eqn\thmR{\eqalign{
I_{A_n}^{(m)}\big (\y;{\tilde \y}^{-1};p,q \big ) = {}&
\prod_{1\leq i,j\leq m+n+4}\!\!\!\!\! \Gamma \big (y_i/{\tilde y}_j\,;p,q \big ) \
I_{A_m}^{(n)}\big (Y^{{1\over m+1}}\y^{-1};{\tilde \y}/{\tilde Y}{}^{{1\over m+1}}\,;p,q \big )
\, , \cr
\hbox{for} \ \
& Y = {\ts {\prod_i}}\, y_i \, , \qquad   {\tilde Y} = {\ts {\prod_i}}\, {\tilde y}_i \, ,
\qquad Y/{\tilde Y}= (p\,q)^{m+1} \, , }
}
implying then exactly \II. Furthermore from \rains
\eqn\thmO{
I_{A_n}^{(0)}\big (\y;{\tilde \y}^{-1}\,;p,q \big ) = 
\prod_{1\leq i,j\leq n+4}\!\!\!\!\! \Gamma \big (y_i/{\tilde y}_j\,;p,q \big ) \
\prod_{1\leq i\leq n+2}\!\!\!
\Gamma \big (Y \y^{-1}, {\tilde \y}/{\tilde Y}\, ;p,q \big ) \, ,
}
with $Y,{\tilde Y}$ as in \thmR. This evaluation of the integral applies when the
magnetic gauge group is trivial.

The detailed expressions in both \indexE\ and \indexM\  depend on the
precise details of the dual gauge groups and assignments of $U(1)_B$ charges for
each theory so this result is a significant test of the details of Seiberg duality 
for these theories. This is in contrast to the large $N_f$, $N_c$ expansions 
of section 5 where many such details were irrelevant. The proof of the theorem in \rains,
see also \proof,
relating these integrals is non trivial and does not involve any straightforward
transformations between each side, it requires demonstrating the result
for particular special cases which are then argued to form a dense set.

\newsec{Indices for Dual Theories with $Sp(2N)$ Gauge Group}

Duality extends to $\N=1$ supersymmetric gauge theories with other
gauge groups. In this section, we consider a gauge group $G=Sp(2N)$,
with a matter sector consisting of $2N_f$ chiral scalar fields $Q$, belonging
to the $2N$ dimensional fundamental representation of the gauge group.
The corresponding flavour symmetry group $F = SU(2N_f)\times U(1)_R $.
The vector multiplet $V$ of course belongs to the $N(2N+1)$ dimensional 
adjoint representation. The overall representation content is summarised in Table 5.

\medskip

\vbox{
\hskip2.5cm Table 5: Electric $Sp(2N)$ Gauge Theory
\nobreak

\hskip2.5cm
\vbox{\tabskip=0pt \offinterlineskip
\hrule
\halign{&\vrule# &\strut \ \hfil#\  \cr
height2pt&\omit&&\omit&&\omit&&\omit&\cr
&\   Field  \ \hfil   && \    $Sp(2N)$  \  && \  $SU(2N_f)$ \
&&  \ $U(1)_R$ \  &\cr
height2pt&\omit&&\omit&&\omit&&\omit&\cr
\noalign{\hrule}
height2pt&\omit&&\omit&&\omit&&\omit& \cr
& \ $Q$  \ \hfil &&  $f$  \hfil     &&    $f$     \hfil           
&&  $1 - (N+1)/ N_f$ \hfil & \cr
& \ $V$  \ \hfil &&  adj.  \hfil  &&  $1$     \hfil         
&&  $1$ \hfil
& \cr
height2pt&\omit&&\omit&&\omit&&\omit& \cr
}
\hrule}
}

The dual theory is a $Sp(2{\wtN})$ gauge theory again, where
\eqn\deftN{
\wtN =N_f-N-2\, ,
}
and with the same flavour symmetry group $F$.
The field content consists of $2N_f$ scalar multiplets $q$, in the $2{\wtN}$
dimensional fundamental representation, a vector multiplet $\tilde V$, in the
$ \wtN (2 \wtN+1)$ dimensional adjoint representation, 
and a gauge singlet scalar multiplet $M$ belonging to the antisymmetric tensor
representation $T_A$ of dimension $N_f(2N_f-1)$ \intrilp.
The representation content is as in Table 6.

\medskip

\vbox{
\hskip2.5cm Table 6: Magnetic $Sp(2\wtN)$ Gauge Theory
\nobreak

\hskip2.5cm
\vbox{\tabskip=0pt \offinterlineskip
\hrule
\halign{&\vrule# &\strut \ \hfil#\  \cr
height2pt&\omit&&\omit&&\omit&&\omit&\cr
&\   Field  \ \hfil   && \    $Sp(2\wtN)$  \  && \  $SU(2N_f)$ \
&&  \ $U(1)_R$ \  &\cr
height2pt&\omit&&\omit&&\omit&&\omit&\cr
\noalign{\hrule}
height2pt&\omit&&\omit&&\omit&&\omit& \cr
& \ $q$  \ \hfil &&  $f$  \hfil     &&    ${\bar f}$    
\hfil            &&  $(N +1)/ N_f$ \hfil & \cr
& \ $\tilde V$  \ \hfil &&  adj.  \hfil     &&    $1$     \hfil     
&&  $1$ \hfil & \cr
& \ $M$  \ \hfil &&  $1$  \hfil     &&    $T_A$     \hfil         
  &&  $2(\wtN+1)/N_f$ \hfil & \cr
height2pt&\omit&&\omit&&\omit&&\omit& \cr
}
\hrule}
}

Imposing $r\ge {1\over 3}$ for both $Q,q$ leads to the conformal window
${3\over 2}(N+1) \le N_f \le 3 (N+1)$.

The single particle index in each case, $i_E(p,q,\y,\z)$ and 
$i_M(p,q,\y,{\tilde \z})$, 
may be straightforwardly formed by applying \ione, using \expforch\ and \cTA\ for 
$SU(2N_f)$ characters $\chi_{SU(2N_f)}(\y)$ with $\y=(y_1, \dots ,y_{2N_f})$. 
The required $Sp(2N)$ and $Sp(2{\wtN})$ characters are obtained from the following 
results for $Sp(2n)$ in general
\eqn\weylsp{\eqalign{
{\chi}_{Sp(2n),f}(\x)={}& \sum_{i=1}^n\big(x_i+x_i{\! }^{-1}\big)\, ,\cr
{\chi}_{Sp(2n),{\rm adj.}}(\x)={}& \! \sum_{1\leq i<j\leq n}\!\!\! \big(x_i\,x_j
+ x_i\,x_j{\!}^{-1}+x_i{\!}^{-1}\, x_j+x_i{\!}^{-1}\,x_j{\!}^{-1}\big)
+\sum_{i=1}^n\big(x_i{\!}^2+x_i{\!}^{-2}\big)+n\, .}
}
For invariant integration over $Sp(2n)$ of any symmetric $f(\x)$ we also have
\eqn\invSp{
\int_{Sp(2n)} \!\!\!\! \d \mu(\x)  \, f(\x) = 
{(-1)^n\over 2^n n!}\int_{\Bbb T_{n}} \prod_{j=1}^{n}
{\d x_j\over 2\pi i x_j} \, \prod_{j=1}^n \big (x_j-x_j{\!}^{-1} \big )^2
\Delta(\x+ \x^{-1})^2  \, f(\x) \, .
}

Assuming the rescaling
\eqn\constraint{
(p\,q)^{(\wtN+1)/2N_f}\y\to \y  \quad \Rightarrow \quad 
\prod_{i=1}^{2N_f}y_i=(p\,q)^{\wtN + 1} \, ,
}
the single particle index then becomes
\eqn\gpoff
{\eqalign{
i_E(p,q,\y,\z)= {}& -\bigg({p\over 1-p}+{q\over 1-q}\bigg)
\, {\chi}_{Sp(2N),{\rm adj.}}(\z) \cr
&{} +{1\over (1-p)(1-q)}\sum_{i=1}^{2N_f}
\big (y_i-p\,q/ y_i \big ) \, {\chi}_{Sp(2N),f}(\z) \,.}
}
As a consequence of \invSp\ the result \itwo\ for the electric index is 
expressible as a multi-contour integral 
\eqn\indexspne{\eqalign{
\!\!\!\! & I_E(p,q,\y)_{Sp(2N)} \cr
&{} ={(-1)^{N}\over 2^{N}N!} \int \prod_{j=1}^{N}
{\d z_j\over 2\pi i z_j} \, \prod_{j=1}^N \big (z_j- z_j{\!}^{-1} \big )^2
\Delta(\z+ \z^{-1})^2 \, 
\exp\bigg(\sum_{n=1}^{\infty}{1\over n}i_E(p^n,q^n,\y^n,\z^n)\bigg) \, .}
}
Using \igauge, \eThe\ and \weylsp, we may write, with the notation \mGam,
\eqn\wotwemay{\eqalign{
&\exp\bigg(-\sum_{n=1}^\infty {1\over n}\Big({p^n\over 1-p^n}+{q^n\over
1-q^n}\Big) \, {\chi}_{Sp(2N),{\rm adj.}}(\z^n) \bigg ) \cr
&={(-1)^{N}(p;p)^{N}\, (q;q)^{N}}{1\over \De \big (\z+\z^{-1}\big )^2
\prod_{1\leq j \leq N}(z_j -z_j{}^{-1})^2}\cr
&\quad\times {1\over \prod_{1\leq i<j\leq N}
\Gamma \big (z_i z_j,z_i/z_j,z_j/z_i,1/(z_iz_j);p,q \big )
\, \prod_{1\leq j \leq N}\Gamma \big (z_j{\!}^2,1/z_j{\!}^2;p,q \big )}\, ,}
}
where the inverse $Sp(2N)$ measure is generated by
\eqn\meas{\eqalign{
& \prod_{1\leq i<j\leq N}(1-z_iz_j)(1-z_i/z_j)(1-z_j/z_i)(1-1/z_iz_j)
= {\Delta ( \z + \z^{-1})^2} \, , \cr
& \prod_{i=1}^N \ (1-z_i{\!}^2 )(1-z_i{}^{-2})
= (-1)^N \prod_{i=1}^N \ (z_i -z_i{}^{-1})^2 \, . }
}
Hence \gpoff\ becomes
\eqn\indexSpE{\eqalign{
& I_E(p,q,\y)_{Sp(2N)} = (p;p)^{N}\, (q;q)^{N} \, {1\over 2^{N}N!}\cr
&\times \int \! \! \prod_{1\leq j\leq N}{\d z_j\over 2\pi i z_j} \, 
{\prod_{1\leq i \leq 2N_f}\prod_{1\leq j \leq N}
\Gamma \big (y_i z_j,y_i/z_j ;p,q \big ) \over \prod_{1\leq i<j\leq N}
\Gamma \big (z_i z_j,z_i/z_j,z_j/z_i,1/(z_iz_j);p,q \big )\, 
\prod_{1\leq j\leq N}\Gamma \big (z_j{\!}^2,1/z_j{\!}^2;p,q \big )} \, ,}
}
where the integrand now involves only elliptic gamma functions
in a similar manner to \indexE. In this case the factors in the integrand denominator 
are associated with the roots for $C_N$, $\pm e_i \pm e_j, \, i \ne j, \, \pm 2e_i$
for $i,j= 1. \dots N$.

For the corresponding magnetic theory the single particle index becomes, using \cTA,
\eqn\important{\eqalign{
i_M(p,q,\y,{\tilde \z})= {}& -\bigg({p\over 1-p}+{q\over 1-q}\bigg)\,
{\chi}_{Sp(2\wtN),{\rm adj.}}({\tilde \z}) \cr
& {} +{1\over (1-p)(1-q)}\bigg((p\,q)^{1\over 2}
\sum_{i=1}^{2N_f} \big (y_i^{-1}-y_i \big )\, {\chi}_{Sp(2{\wtN}),f}({\tilde \z})\cr
\noalign{\vskip -6pt}
&\hskip 3cm{} +\sum_{1\leq i< j\leq 2 N_f}\big (y_iy_j -p\,q/(y_iy_j) \big )\bigg) \, . }
}
The magnetic index is then
\eqn\indexspm{\eqalign{
I_M(p,q,\y)_{Sp(2\wtN)} = {}& {(-1)^{\wtN}\over 2^{\wtN}\wtN!} \int \prod_{j=1}^{\wtN}
{\d \tz_j\over 2\pi i \tz_j} \, \prod_{j=1}^{\wtN} \big (\tz_j- \tz_j{\!}^{-1} \big )^2
\Delta({\tilde \z} + {\tilde z}^{-1})^2 \cr
\noalign{\vskip -8pt}
& \hskip 3.2cm {}\times 
\exp\bigg(\sum_{n=1}^{\infty}{1\over n}i_M(p^n,q^n,\y^n,{\tilde \z}^n)\bigg) \, .}
}
and, in the same fashion as \indexSpE\ was obtained, we now have
\eqnn\indexSpM
$$\eqalignno{
\!\!\! & I_M(p,q,\y)_{Sp(2\wtN)} 
= \prod_{1\leq i<j\leq 2N_f}\!\!\!  \Gamma \big (y_i\,y_j;p,q \big ) \ 
(p;p)^{\wtN}\, (q;q)^{\wtN} \, {1\over 2^{\wtN}\wtN!}\cr
\!\!\! &\times \int \! \! \prod_{1\leq j\leq \wtN}{\d \tz_j\over 2\pi i \tz_j} \, 
{\prod_{1\leq i \leq 2N_f}\prod_{1\leq j\leq \wtN}
\Gamma \big (t \tz_j/y_i, t/(y_i \tz_j) ;p,q \big )\over 
\prod_{1\leq i<j\leq \wtN}
\Gamma \big (\tz_i \tz_j,\tz_i/\tz_j,\tz_j/\tz_i,1/(\tz_i\tz_j);p,q \big )\, 
\prod_{1\leq j\leq \wtN}\Gamma \big (\tz_j{}^2,1/\tz_j{}^2;p,q \big )} \cr
&{} = \prod_{1\leq i<j\leq 2N_f}\!\!\!  \Gamma \big (y_i\,y_j;p,q \big ) \ 
I_E(p,q,t\y^{-1})_{Sp(2\wtN)} \, , & \indexSpM }
$$
for $t= (p\,q)^{1\over 2}$.

Again, happily, the relevant integrals were considered by Rains \rains. The 
right hand side of \indexSpE\ for $n= N, \, m= {\tilde N}$ defines the 
elliptic hypergeometric integral
\eqn\BCn{\eqalign{
&  I_{BC_n}^{(m)}\big ( \y ; p,q\big ) = (p;p)^{n}\, (q;q)^{n} \, {1\over 2^{n}n!}\cr
&\times \int \! \! \prod_{1\leq j\leq n}{\d z_j\over 2\pi i z_j} \,
{\prod_{1\leq i \leq 2(m+n+2)}\prod_{1\leq j \leq n}
\Gamma \big (y_i z_j,y_i/z_j ;p,q \big ) \over \prod_{1\leq i<j\leq n}
\Gamma \big (z_i z_j,z_i/z_j,z_j/z_i,1/(z_iz_j);p,q \big )\,
\prod_{1\leq j\leq n}\Gamma \big (z_j{\!}^2,1/z_j{\!}^2;p,q \big )} \, ,}
}
depending on a ${2(m+n+2)}$-dimensional vector $\y$.
Theorem 3.1 of \rains\ requires
\eqn\thmB{
I_{BC_n}^{(m)}\big ( \y\, ; p,q\big ) = 
\prod_{1\leq i<j\leq 2(m+n+2)}\!\!\!\!\!\!\! \Gamma \big (y_i\,y_j;p,q \big ) \
I_{BC_m}^{(n)}\big ( \sqrt{pq}\ \y^{-1}; p,q\big ) \, , \ \hbox{for} \ 
{\ts {\prod_i}} \, y_i  = (p\,q)^{m+1} \, .
}
This then implies $I_E(p,q,\y)_{Sp(2N)}$ 
and $I_M(p,q,\y)_{Sp(\wtN)}$ in \indexSpE\ and \indexSpM\ are equal. 
In this case $I_{BC_0}^{(m)}\big ( \y ; p,q\big )=1$. Applying the transformation
twice leads to the identity as a consequence of \GG.

\newsec{Indices for Dual Theories with $SO(N)$ Gauge Groups}

The original paper on duality \Sei, see also \IntS, discussed additionally
$\N=1$ theories with orthogonal
gauge groups with $N_f$ chiral quark fields in the vector representation, so that
the flavour symmetry group  $F = SU(N_f)\times U(1)_R $. The adjoint 
representation here has dimension $\half N(N-1)$.
The overall representation content is summarised in Table 7.

\medskip

\vbox{
\hskip2.5cm Table 7: Electric $SO(N)$ Gauge Theory
\nobreak

\hskip2.5cm
\vbox{\tabskip=0pt \offinterlineskip
\hrule
\halign{&\vrule# &\strut \ \hfil#\  \cr
height2pt&\omit&&\omit&&\omit&&\omit&\cr
&\   Field  \ \hfil   && \    $SO(N)$  \  && \  $SU(N_f)$ \
&&  \ $U(1)_R$ \  &\cr
height2pt&\omit&&\omit&&\omit&&\omit&\cr
\noalign{\hrule}
height2pt&\omit&&\omit&&\omit&&\omit& \cr
& \ $Q$  \ \hfil &&  vec.  \hfil     &&    $f$     \hfil           
&&  $1 - (N-2)/ N_f$ \hfil & \cr
& \ $V$  \ \hfil &&  adj.  \hfil  &&  $1$     \hfil         
&&  $1$ \hfil
& \cr
height2pt&\omit&&\omit&&\omit&&\omit& \cr
}
\hrule}
}

The dual theory is also a $SO({\wtN})$ gauge theory, where
\eqn\deftO{
\wtN =N_f-N + 4 \, ,
}
and with the same flavour symmetry group $F$.
The field content consists of $N_f$ scalar multiplets $q$, in the
vector representation, a vector multiplet $\tilde V$, in the
$ \half \wtN (\wtN- 1)$ dimensional adjoint representation, 
and a gauge singlet scalar multiplet $M$ belonging to the symmetric tensor
representation $T_S$ of dimension $\half N_f(N_f+1)$.
The representation content is as in Table 8.

\medskip

\vbox{
\hskip2.5cm Table 8: Magnetic $SO(\wtN)$ Gauge Theory
\nobreak

\hskip2.5cm
\vbox{\tabskip=0pt \offinterlineskip
\hrule
\halign{&\vrule# &\strut \ \hfil#\  \cr
height2pt&\omit&&\omit&&\omit&&\omit&\cr
&\   Field  \ \hfil   && \    $SO(\wtN)$  \  && \  $SU(N_f)$ \
&&  \ $U(1)_R$ \  &\cr
height2pt&\omit&&\omit&&\omit&&\omit&\cr
\noalign{\hrule}
height2pt&\omit&&\omit&&\omit&&\omit& \cr
& \ $q$  \ \hfil &&  vec.  \hfil     &&    ${\bar f}$    
\hfil            &&  $(N-2)/ N_f$ \hfil & \cr
& \ $\tilde V$  \ \hfil &&  adj.  \hfil     &&    $1$     \hfil     
&&  $1$ \hfil & \cr
& \ $M$  \ \hfil &&  $1$  \hfil     &&    $T_S$     \hfil         
&&  $2 - 2 (N-2)/N_f$ \hfil & \cr
height2pt&\omit&&\omit&&\omit&&\omit& \cr
}
\hrule}
}

Imposing $r\ge {1\over 3}$ for both $Q,q$ leads to the conformal window
${3\over 2}(N-2) \le N_f \le 3(N-2) $.

For characters for $SO(N)$ it is necessary to distinguish according to whether
$N$ is even or odd. For $\x=(x_1,\dots , x_n)$ the relevant results are
\eqn\weylev{\eqalign{
{\chi}_{SO(2n),{\rm vec.}}(\x)={}& \sum_{i=1}^n\big(x_i+x_i{\! }^{-1}\big)\, ,\cr
{\chi}_{SO(2n),{\rm adj.}}(\x)={}& \! \sum_{1\leq i<j\leq n}\!\!\! \big(x_i\,x_j
+ x_i\,x_j{\!}^{-1}+x_i{\!}^{-1}\, x_j+x_i{\!}^{-1}\,x_j{\!}^{-1}\big) +n\, ,}
}
and
\eqn\weylod{\eqalign{
{\chi}_{SO(2n+1),{\rm vec.}}(\x)={}& 
\sum_{i=1}^n\big(x_i+x_i{\! }^{-1}\big) + 1 \, ,\cr
{\chi}_{SO(2n+1),{\rm adj.}}(\x)={}& \! 
\sum_{1\leq i<j\leq n}\!\!\! \big(x_i\,x_j
+ x_i\,x_j{\!}^{-1}+x_i{\!}^{-1}\, x_j+x_i{\!}^{-1}\,x_j{\!}^{-1}\big)
+\sum_{i=1}^n\big(x_i+x_i{\! }^{-1} \big)+n\, .}
}
We also require
\eqn\cTS{
\chi_{SU(n),T_S}^{\vphantom g} (\x)  = \sum_{1\leq i<j \leq n} \!\!\! x_i x_j 
+ \sum_{i=1}^n x_i{\!}^2 \, , \qquad
\chi_{SU(n),{\bar T}_S}^{\vphantom g}(\x ) = 
\chi_{SU(n),T_S}^{\vphantom g} (\x^{-1})\, .
}
For invariant integration over $SO(N)$ of any symmetric $f(\x)$ we also have
\eqn\invSO{\eqalign{
\int_{SO(2n)} \!\!\!\! \d \mu(\x)  \, f(\x) = {}& 
{1 \over 2^{n-1} n!}\int_{\Bbb T_{n}} \prod_{j=1}^{n}
{\d x_j\over 2\pi i x_j} \,  \Delta(\x+ \x^{-1})^2  \, f(\x) \, ,\cr
\int_{SO(2n+1)} \!\!\!\!\!\!\! \d \mu(\x)  \, f(\x) = {}&
{(-1)^n\over 2^n n!}\int_{\Bbb T_{n}} \prod_{j=1}^{n}
{\d x_j\over 2\pi i x_j} \,  \prod_{j=1}^n \big 
( x_j{\!}^{{1\over 2}} - x_j{\!}^{-{1\over2}} \big )^2 \, 
\Delta(\x+ \x^{-1})^2 \, \, f(\x) \, .\cr }
}

The single particle indices \ione\ are obtained in a similar fashion as previously 
\eqn\iEO
{\eqalign{
i_E(p,q,\y,\z)= {}& -\bigg({p\over 1-p}+{q\over 1-q}\bigg)
\, {\chi}_{SO(N),{\rm adj.}}(\z) \cr
&{} +{1\over (1-p)(1-q)}\sum_{i=1}^{N_f}
\big (y_i-p\,q/ y_i \big ) \, {\chi}_{SO(N),{\rm vec.}}(\z) \, , }
}
and
\eqn\iMO{\eqalign{
\!\!\!\! i_M(p,q,\y,{\tilde \z})= {}& -\bigg({p\over 1-p}+{q\over 1-q}\bigg)\,
{\chi}_{SO(\wtN),{\rm adj.}}({\tilde \z}) \cr
& {} +{1\over (1-p)(1-q)}\bigg((p\,q)^{1\over 2}
\sum_{i=1}^{N_f} \big (y_i^{-1}-y_i \big )\,
{\chi}_{SO({\wtN}),{\rm vec.}}({\tilde \z})\cr
\noalign{\vskip -6pt}
&\hskip 2.5cm{} + \!\!\!
\sum_{1\leq i< j\leq N_f}\!\!\!\! \big (y_iy_j -p\,q/(y_iy_j) \big )
+ \sum_{i=1}^{N_f}\big (y_i{\!}^2 -p\,q/y_i{\!}^2 \big ) \bigg) \, , }
}
where $\y$ has been rescaled so that
\eqn\cons{
\prod_{i=1}^{N_f}y_i=(p\,q)^{{1\over 2}(N_f-N+2) } \, .
}

The integral formulae for the index are then generated very much as before. 
The adjoint characters in \weylev\ and \weylod\ generate contributions
which cancel the integration measures in \invSO\ by using \meas\ once more.
Hence, taking $N=2n$ and $N=2n+1$,
\eqn\indexOEe{\eqalign{
I_E(p,q,\y)_{SO(2n)}{}&  = (p;p)^{n}\, (q;q)^{n} \, {1\over 2^{n-1}n!}\cr
&\times \int \! \! \prod_{1\leq j\leq n}{\d z_j\over 2\pi i z_j} \, 
{\prod_{1\leq i \leq N_f}\prod_{1\leq j \leq n}
\Gamma \big (y_i z_j,y_i/z_j ;p,q \big ) \over \prod_{1\leq i<j\leq n}
\Gamma \big (z_i z_j,z_i/z_j,z_j/z_i,1/(z_iz_j);p,q \big )\, } \, ,}
}
and
\eqn\indexOEo{\eqalign{
& I_E(p,q,\y)_{SO(2n+1)} = 
(p;p)^{n}\, (q;q)^{n} \,
\prod_{1\leq i\leq N_F} \Gamma(y_i;p,q) \ {1\over 2^{n}n!} \cr
&\times \int \! \! \prod_{1\leq j\leq n}{\d z_j\over 2\pi i z_j} \, 
{\prod_{1\leq i \leq N_f}\prod_{1\leq j \leq n}
\Gamma \big (y_i z_j,y_i/z_j ;p,q \big ) \over \prod_{1\leq i<j\leq n}
\Gamma \big (z_i z_j,z_i/z_j,z_j/z_i,1/(z_iz_j);p,q \big )\, 
\prod_{1\leq j\leq n}\Gamma \big (z_j,1/z_j;p,q \big )} \, .}
}
In \indexOEe\ and \indexOEo\ the factors in the integrand denominator 
may be matched with the roots
for $D_N$, $\pm e_i \pm e_j, \,  i<j$, and $B_N$, $\pm e_i \pm e_j , \, 
i<j, \ \pm e_j$, respectively.

For the corresponding magnetic theory  the results are very similar except
for contributions involving the meson field $M$, which are obtained from
the last line of \iMO. The results are expressed concisely as 
\eqn\indexOMe{
I_M(p,q,\y)_{\smash{SO(\wtN)}} = 
\prod_{1\leq i<j\leq N_f}\!\! \Gamma \big (y_i\,y_j;p,q \big )\!
\prod_{1\leq i \leq N_f}\!\! \Gamma \big (y_i{\!}^2 ;p,q \big )\, 
I_E(p,q,\sqrt{pq}\, \y^{-1})_{\smash{SO(\wtN)}}  \, .
}
The required identity is then
\eqn\finO{
I_E(p,q,\y)_{SO(N)} = I_M(p,q,\y)_{\smash{SO(\wtN)}} \, , 
\ \hbox{for} \ {\ts {\prod}_i} y_i = (pq)^{{1\over 2}\wtN-1} \, , \
N_f = N + \wtN - 4 \, .
}

The relation \finO\ involving
$B_N$ and $D_N$ multi-variable elliptic beta integrals can be reduced
to a special case of \thmB\ by virtue of an argument due to Rains \rainsB.
It is easy to verify
\eqn\Gamsq{
\Gamma(z^2; p,q ) = {\ts {\prod_a}} \Gamma(z u_a ; p,q ) \, , \qquad
\u = \big ( 1,-1,p^{1\over 2}, - p^{1\over 2}, q^{1\over 2}, -  q^{1\over 2},
(pq)^{1\over 2}, - (pq)^{1\over 2} \big ) \, ,
}
With the definition \indexOEe\ we may then express the index in terms of
$I_{BC_n}^{(m)}$ as in \BCn\ by
\eqn\Irel{
I_E(p,q,\y)_{SO(2n)} = \cases{ 2\, I_{BC_n}^{({1\over 2}(N_f+4)-n)}(\y,\u;p,q) \, , 
&$N_f$ even; \cr
2\, I_{BC_n}^{({1\over 2}(N_f+3) -n)}(\y,\v;p,q) \, , &$N_f$ odd; }
}
where, noting that $\Gamma((pq)^{1\over 2} z_j , (pq)^{1\over 2}/ z_j ; p,q) =1$
as a consequence of \GG, we also define
\eqn\defv{
\v = \big ( 1,-1,p^{1\over 2}, - p^{1\over 2}, q^{1\over 2}, -  q^{1\over 2},
- (pq)^{1\over 2} \big ) \, . 
}
In a similar vein starting from \indexOEo\ we may also write
\eqn\Irelo{
I_E(p,q,\y)_{SO(2n+1)} = \cases{ {\ts{\prod_{i=1}^{N_F}}}\, \Gamma(y_i;p,q)\
I_{BC_n}^{({1\over 2}(N_f+4)-n)}(\y,\u';p,q) \, , 
&$N_f$ even; \cr
{\ts{\prod_{i=1}^{N_F}}}\, \Gamma(y_i;p,q)\ 
I_{BC_n}^{({1\over 2}(N_f+5) -n)}(\y,\v';p,q) \, , &$N_f$ odd; }
}
for
\eqn\uvp{\eqalign{
\u' = {}&\big ( -1,p^{1\over 2}, - p^{1\over 2}, q^{1\over 2}, -  q^{1\over 2},
- (pq)^{1\over 2} \big ) \, , \cr
\v' = {}& \big (-1,p^{1\over 2}, - p^{1\over 2}, q^{1\over 2}, -  q^{1\over 2},
- (pq)^{1\over 2} , (pq)^{1\over 2} \big ) \, . \cr}
}
The necessary identity to ensure \indexOMe\ and \finO\ then follows from \thmB,
taking into account $\v' \sim \sqrt{pq} \, \v^{-1}$ and the results
\eqnn\resG
$$\eqalignno{
{\ts{\prod_{a}}} \Gamma( y_i u_a ; p,q ) ={}&  \Gamma ( y_i{\!}^2; p, q) \, , \qquad
{\ts{\prod_{a}}} \Gamma( y_i v_a ; p,q ) = \Gamma ( y_i{\!}^2; p, q)\,
\Gamma ( \sqrt{pq}/ y_i ; p, q) \, , \cr
{\ts{\prod_{a}}} \Gamma( y_i u'{\!}_a ; p,q ) ={}&  \Gamma ( y_i{\!}^2; p, q) \, 
{ \Gamma ( \sqrt{pq}/ y_i ; p, q)  \over \Gamma ( y_i; p, q) } \, , \qquad
{\ts{\prod_{a}}} \Gamma( y_i v'{\!}_a ; p,q ) = {\Gamma ( y_i{\!}^2; p, q)
\over \Gamma ( y_i; p, q) }  \, , \cr
{\ts{\prod_{a<b}}} \Gamma( u_a u_b ; p,q )  = {}& 
{\ts{\prod_{a<b}}} \Gamma( u'{\!}_a u'{\!}_b ; p,q ) = 1 \, , \cr
2 \, {\ts{\prod_{a<b}}} \Gamma( v_a v_b ; p,q )  = {}&  
\half \, {\ts{\prod_{a<b}}} \Gamma( v'{\!}_a v'{\!}_b ; p,q ) = 1 \, . &\resG
}
$$

We also test the result in the simple case $N=4, \, N_f=3, \,
{\tilde N}=3$ which involves duality between $SO(N)$ gauge theories
with even and odd $N$ by considering the first few terms in an expansion.
As a result of $SO(4) = SU(2)\times SU(2)/\Bbb Z_2$
and $SO(3) = SU(2)/\Bbb Z_2$ we have, letting for $SO(4)$ $z_1 = uv, \, z_2 = u/v$
and for $SO(3)$ $z_1 = w^2$, from \invSO
\eqn\intSU{
\int_{SO(4)} \!\!\! \d \mu(\z) = \int_{SU(2)} \!\!\!\!\! \d \mu(v)\,
\int_{SU(2)} \!\!\!\!\! \d \mu(u)\, ,\qquad
\int_{SO(3)} \!\!\! \d \mu(\z) = \int_{SU(2)} \!\!\!\!\! \d \mu(w)\, ,
}
since $\De(\z+\z^{-1})^2 = (1-u^2)^2(1-v^2)^2/u^2v^2$. With $p=tx, \, q=tx^{-1}$
the electric single particle index from \iEO\ becomes
\eqn\speE{\eqalign{
i_E(t x,t x^{-1},\y,u,v) = {}& {1\over (1-tx)(1-tx^{-1})} \Big (
\big ( 2t^2 - t \chi_2(x) \big ) \big ( \chi_3 (u) + \chi_3 (v) \big)\cr
\noalign{\vskip -6pt}
&\hskip 3cm{}+ 
\big ( p_3(\y) - t^2 p_3(\y^{-1}) \big ) \chi_2(u) \chi_2(v) \Big ) \, , }
}
for 
\eqn\consp{
\y=(y_1,y_2,y_3) \, , \qquad y_1 y_2 y_3 = t \, ,
} 
and with $\chi_2,\chi_3$ defined in \chit. For the magnetic index from \iMO,
\eqn\speM{\eqalign{
i_M(t x,t x^{-1},\y,w) = {}& {1\over (1-tx)(1-tx^{-1})} \Big (
\big ( 2t^2 - t \chi_2(x) + t \, p_3(\y^{-1}) - t \, p_3(y) \big )
\chi_3 ( w ) \cr
\noalign{\vskip -6pt}
&\hskip 3cm{}+ s_{(2,0)}(\y) - t^2 s_{(2,0)} (\y^{-1}) \Big ) \, , \cr
& s_{(2,0)}(\y) = \half \big ( p_3(y)^2 + p_3 (\y^2) \big ) \, . }
}
The required index identity is then from \itwo
\eqn\idind{\eqalign{
I(t x,t x^{-1},\y) = {}& \int_{SU(2)} \!\!\!\!\! \d \mu(v)\,
\int_{SU(2)} \!\!\!\!\! \d \mu(u)\,  \exp \bigg ( \sum_{n=1}^\infty {1\over n} \,
i_E \big (t^n ,x^n, \y^n , u^n , v^n \big ) \bigg ) \cr
= {}& \int_{SU(2)} \!\!\!\!\! \d \mu(w)\, \exp \bigg ( \sum_{n=1}^\infty {1\over n} \,
i_M \big (t^n ,x^n, \y^n , w^n \big ) \bigg ) \, .}
}

It is straightforward to expand \idind\ where we may use 
$\chi_2(u^n) = \chi_{n+1}(u) - \chi_{n-1}(u)$, 
$\chi_3(u^n) = \chi_{2n+1}(u) - \chi_{2n-1}(u) + 1$ and apply  standard $SU(2)$
tensor product rules to decompose products of $\chi_n$ into single characters.
The $SU(2)$ integrals can then be evaluated using orthonormality of 
characters or equivalently just by evaluating residues. The index has an expansion
\eqn\exI{
I(t x,t x^{-1},\y) = 1 +  \sum_{n>0} f_n(t,x,\y) \, , \qquad 
f_n(t,x,\y) = {\rm O}\big (t^{{1\over 3}n} \big )\, ,
}
where from \consp\ 
$\y= {\rm O}(t^{{1\over 3}})$. We have checked that both the electric and magnetic
contributions to \idind\ are the same up to ${\rm O}(t^4)$ 
and give the following non zero terms, in terms of $SU(3)$ Schur polynomials 
$s_{(\lambda,\mu)}(\y)$ and $SU(2)$ characters $\chi_{2j+1}(x)$,
\eqn\testI{\eqalign{
f_2(t,x,\y) = {}&  s_{(2,0)}(\y) \, , \cr
f_4(t,x,\y) = {}& s_{(4,0)}(\y)  + s_{(2,2)}(\y) \, ,\cr
f_5(t,x,\y) = {}& t \chi_2(x) \big ( s_{(2,0)}(\y) - s_{(1,1)}(\y) \big ) \, , \cr
f_6(t,x,\y) = {}& s_{(6,0)}(\y) + s_{(4,2)}(\y) - t \, s_{(2,1)}(\y) + 2t^2  \, , \cr
f_7(t,x,\y) = {}& t\, \chi_2(x) \big ( s_{(4,0)}(\y) + s_{(2,2)}(\y) \big ) \, , \cr
f_8(t,x,\y) = {}& s_{(8,0)}(\y) + s_{(6,2)}(\y) + s_{(4,4)}(\y)
- t \, s_{(4,1)}(\y) - t\, s_{(3,2)}(\y) - t^2 s_{(1,1)}(\y)\cr
&{} + t^2 \chi_3(x) \big ( s_{(2,0)}(\y) - s_{(1,1)}(\y) \big ) \, , \cr
f_9(t,x,\y) = {}& t\, \chi_2(x) \big ( s_{(6,0)}(\y) + 2 s_{(4,2)}(\y) - s_{(3,3)}(\y) 
+ t\, s_{(2,1)}(\y) + t^2  \big ) \, , \cr
f_{10}(t,x,\y) = {}& s_{(10,0)}(\y) + s_{(8,2)}(\y) + s_{(6,4)}(\y) \cr
&{} -t s_{(6,1)}(\y) -t s_{(5,2)}(\y) - t s_{(4,3)}(\y) - 2 t^2 s_{(3,1)}(\y)   \cr
&{} + t^2 \chi_3(x) \big ( 2 s_{(4,0)}(\y) - s_{(3,1)}(\y) +2 s_{(2,2)}(\y) - 
t\, s_{(1,0)}(\y) \big ) \, , \cr
f_{11}(t,x,\y) = {}& t\, \chi_2(x) \big ( s_{(8,0)}(\y) 
+ 2 s_{(6,2)}(\y) +  s_{(4,4)}(\y) + t^2 s_{(2,0)}(\y) - t^2 s_{(1,1)}(\y) \big ) \cr
&{} + t^3 \chi_4(x) \big ( s_{(2,0)}(\y) - s_{(1,1)}(\y) \big ) \, , \cr
f_{12}(t,x,\y) = {}& s_{(12,0)}(\y) + s_{(10,2)}(\y) + s_{(8,4)}(\y) + s_{(6,6)}(\y) \cr
&{} -t\, s_{(8,1)}(\y) -t\, s_{(7,2)}(\y) - t\, s_{(6,3)}(\y)  - t\, s_{(5,4)}(\y) \cr
&{}  - 2 t^2 s_{(5,1)}(\y)  -  t^2 s_{(4,2)}(\y) +  2 t^2 s_{(3,3)}(\y) 
+  2 t^3 s_{(3,0)}(\y) + t^3 s_{(2,1)}(\y) - t^4 \cr
&{} + t^2 \chi_3(x) \big ( 2 s_{(6,0)}(\y) + 3 s_{(4,2)}(\y)  -  2 s_{(3,3)}(\y) - 
t\, s_{(3,0)}(\y) + t\, s_{(2,1)}(\y) + 2 t^2 \big ) \, . \cr
}
}
These results are sensitive to all terms which are in $i_E$ and $i_M$ in \speE\ and
\speM, and therefore provide good support for the required all orders result \idind.
It is significant to note also that all coefficients are integers in accord with
the expectation in \expI.

\newsec{Conclusions}

This paper has demonstrated that the naive prescription for the superconformal index 
given by \ione\ and \itwo\ and using the standard results for dual $\N=1$ gauge
theories, where the matter content and its $R$-charges are determined by
careful matching of the spectrum of gauge invariant operators and also
matching the 't Hooft anomalies, leads to results which are the same in both
dual theories. The exact equality of the two expressions for the index has been shown for 
theories in which there is no superpotential and then depends on very non-trivial 
$q$-series type integral identities, only recently proved, which are only valid for the 
detailed
$R$-charges and gauge groups determined by the consistency conditions for duality.
This remarkable correspondence perhaps lends credence to the results for the index
described here following on from  R\"omelsberger \romel.  The elliptic hypergeometric
functions which are generated by the index, and whose non trivial transformation 
properties are a necessary requirement for duality, are also relevant to other areas 
such as quantum integrable systems, \spit.

The situation when there is a superpotential, as in the  Kutasov-Schwimmer case, is less 
clear. The operator spectrum is then constrained by equations of motion and the result 
for the index should be modified.
Nevertheless we also verified that the naive formula for the index gave results which
agreed in the large $N$ limit and also showed that the leading finite $N$ correction was
also consistent.  Perhaps physical considerations may suggest novel identities
which have not yet been proved. Seiberg duality has been extended to a much
wider class of $\N=1$ theories than those considered in this paper, including theories
with exceptional gauge groups \Except.

A remaining issue concerns the precise derivation of the formula for the index
provided by applying \ione\ and \itwo. In particular other than in the free case
when $r={2\over 3}$ the results for chiral fields given by \ione\ have not been
derived in this paper. For interacting theories it is necessary to consider
the superconformal algebra in \QSfree\ and \QVfree\ with in general $F$ and $\DD$
non zero. However, letting for instance $F \to {\bar \vphi}^n$ for some $n$
still enforces $r={2\over 3}$ as a consequence of the commutator $[S^\alpha,F]$.
Similar considerations apply for other modifications although the derivatives 
in \QbarS\ and \QbarV\ may be replaced by gauge covariant derivatives by allowing
for the algebra to be extended by  appropriate gauge transformations.  Perhaps
further inclusion of internal symmetry transformations is necessary at non trivial
superconformal fixed points. This is perhaps suggested by the rescaling of internal
symmetry character variables, such as in \redefvaro, which was a necessary feature
of the analysis of the integrals defining the index at the interacting $\N=1$
superconformal fixed points, at which the duality between electric and magnetic
theories that is considered in this paper is fully realised.

\medskip
\noindent
{\bf Acknowledgements}

We are grateful to Kenneth Intriligator for useful e-mail
correspondence and to Christian R\"omelsberger for
helpful discussions.    F.A.D. is very grateful to the
University of Southampton for hospitality during
completion of this paper.  This work is supported by
an IRCSET (Irish Research Council in Science, Engineering and
Technology) research fellowship. H.O. thanks the Moore
Library for rapidly acquiring the book in \spi. We would also like
to thank Eric Rains for providing the argument for the $SO(N)$ index.
\vfill\eject

\appendix{A}{$\N=1$ Superconformal Representation Theory and Characters}

Using the notation of \char, the generators of the $\N=1$ superconformal group 
$SU(2,2|1)$ consist of those for Lorentz transformations $M_{a b}$, translations
$P_a$, special conformal transformations $K_a$, $a,b=1,\dots, 4$, dilatations
$H$, which is the Hamiltonian in radial quantisation of conformal theories,
the $U(1)_R$ $R$-charge $R$ along with supercharges
$Q_{\al},\, {\bar Q}_{\smash{\dal}}$ and their superconformal partners
$S^{\al},\,{\bar S}^{\dal}$, $\al,\dal=1,2$.  
In a spinorial basis $P_{\alpha\dal} = (\si^a)_{\alpha\dal} P_a$,
${K}{}^{\dal\alpha} = (\bsi^a)^{\dal\alpha} K_a$,
$ M_\alpha{}^{\! \beta} = - \quar i ( \si^a \bsi^b)_\alpha{}^{\! \beta}
M_{ab}$, ${\bar M}{}^\dal{}_{\!\smash{\dbe}} = - \quar i
( \bsi^a \si^b)^\dal{}_{\!\smash{\dbe}} M_{ab}${\foot{The standard hermiticity 
requirements are
$$
\left( \M_\A{}^{\! \B}  \right)^\dagger = (\tau \M \tau)_\B{}^{\! \A} \, ,
\quad R^\dagger = R \, , \quad
\left(\Q_\A\right)^\dagger = ({\bar \Q}\tau)^\A \, , \quad \tau =
\pmatrix{0&1\cr 1&0} \, .
$$
Thus $H^\dagger =-H$ and $(M_\alpha{}^{\! \beta})^\dagger =
{\bar M}{}^\dbe{}_{\!\smash{\dal}}$, interchanging $SU(2)_J$ and
$SU(2)_\bJ$.}
}. With the notation,
\eqn\defMab{
\M_\A{}^{\! \B} = \pmatrix{ M_\alpha{}^{\! \beta} + \half
\de_\alpha{}^{\! \beta} H & \half \, P_{\smash{\alpha\dbe}} \cr
\half \, {K}{}^{\dal\beta} & {\bar M}{}^\dal{}_{\!\smash{\dbe}}
- \half \de{}^\dal{}_{\!\smash{\dbe}} H \cr} \, , \ \
\Q_\A = \pmatrix{Q_{ \alpha}\cr  \bS^{\dal}} \, , \ \
{\bar \Q}^\B = \pmatrix{ S^{\!\beta}& \bQ_{\smash {\dbe}} }\, ,
}
the $SU(2,2|1)$ algebra is expressible as
\eqn\comm{\eqalign{
\big [ \M_\A{}^{\! \B}, \M_\C{}^{\! \D} \big ] = {}& \de_\C{}^{\! \B}
\M_\A{}^{\! \D} - \de_\A{}^{\! \D} \M_\C{}^{\! \B} \, , \cr
\big [ \M_\A{}^{\! \B}, \Q_\C \big ] = {}& \de_\C{}^{\! \B}
\Q_\A - \quar \de_\A{}^{\! \B} \Q_\C \, , \qquad
\big [ \M_\A{}^{\! \B}, {\bar \Q}^\C \big ] = - \de_\A{}^{\! \C}
{\bar \Q}^\B + \quar \de_\A{}^{\! \B} {\bar \Q}^\C \, , \cr
\big [ R ,  \Q_\A \big ] = {}& -\Q_\A \, , \qquad \,
\big [R , {\bar \Q}^\B \big ] =  {\bar \Q}^\B\, , \cr
\big \{ \Q_\A , {\bar \Q}^\B \big \} = {}& 4
\M_\A{}^{\! \B} +3 \de_\A{}^{\! \B} R\, , \quad
\big \{ \Q_\A , \Q_\B  \big \} = 0 \, , \quad
\{ {\bar \Q}^\A , {\bar \Q}^\B \big \} = 0 \, ,\cr}
}
for $\de_\A{}^{\! \B} = \pmatrix{\de_\alpha{}^{\! \beta}&0\cr 0&
\de{}^\dal{}_{\!\smash{\dbe}}}$.
In terms of the usual angular momentum generators we have,
\eqn\MJ{
\big [M_\alpha{}^{\! \beta}\big ] = \pmatrix{ J_3 & J_+ \cr J_- & -J_3 } \, ,
\qquad \big [{\bar M}{}^\dbe{}_{\!\smash{\dal}}\big ] =
\pmatrix{ \bJ_3 & \bJ_+ \cr \bJ_- & -\bJ_3 } \, ,
}
with $[J_+,J_-]=2J_3$, $[\bJ_+,\bJ_-]=2\bJ_3$.

A generic highest weight primary state for this superalgebra
$|\De,r,j,\bj\rangle^{\rm h.w.}$, which has conformal
dimension $\De$, belongs to the spin $SU(2)_J\times SU(2)_\bJ$ representation 
$(j,\bj)$ and has $R$-symmetry eigenvalue $r$, satisfies
\eqn\hws{\eqalign{
(K^{\dal\al},S^{\al},{\bar S}^{\dal},J_{+},{\bar J}_+)
|\De,r,j,\bj\rangle^{\rm h.w.}& {}=0\, , \cr
 (H, R,J_3,{\bar J}_3) |\De,r,j,\bj\rangle^{\rm h.w.}
&{}= (\De, r, j , \bj ) |\De,r,j,\bj\rangle^{\rm h.w.} \, . \cr}
}
The corresponding Verma module $\V_{(\De,r,j,\bj)}$ is then
spanned by the states
\eqn\vermam{
\prod_{
\al,\dal,\beta,\dbe=1,2}(P_{\al\dal})^{N_{\al\dal}}
(Q_{\beta})^{n_{\beta}}
({\bar Q}_{\dbe})^{{\bar n}_{\dbe}}(J_-)^{N}({\bar J}_-)^{{\bar N}}
|\De,r,j,\bj\rangle^{\rm h.w.}\, , }
for $N_{\al\dal},N,\,{\bar N}, = 0,1,2, \dots$ and
$n_{\beta}^{\vphantom g},\,{\bar n}_{\smash{\dbe}}^{\vphantom g} =0,1$.

When BPS conditions involving different supercharges are imposed there 
are truncated Verma modules and $\Delta$ is determined in terms of $r,j,\bj$,
although there may also be various other potential constraints on $r,j,\bj$.
For unitary representations the following conditions are relevant,
labelled by $t,{\bar t}$ according to  the fraction of the 
$Q,{\bar Q}$ supercharges to be omitted from \vermam,
\eqna\semis
$$\eqalignno{
& {\bar t}=\half: \Delta = 2+2\bj+{\ts{3\over 2}}r \, , \cr
\noalign{\vskip-3pt}
& \Big ( {\bar Q}_{1} + {1\over 2\bj }\,   {\bar Q}_{2} \bJ_- \Big )
|\De,r,j,\bj\rangle^{\rm h.w.} = 0 \, , \  \bj > 0 \, ,\qquad 
\bQ^2 |\De,r,j, 0 \rangle^{\rm h.w.} = 0 \, , & \semis{a} \cr
& {t}=\half: \Delta = 2+2j-{\ts{3\over 2}}r\, , \cr
\noalign{\vskip-3pt}
& \Big ( Q_2 - {1\over 2j}\, Q_1 J_- \Big )
|\De,r,j,\bj\rangle^{\rm h.w.}  = 0 \, , \ j >0 \, ,\qquad
Q^2 |\De,r,0,\bj\rangle^{\rm h.w.} = 0 \, , & \semis{b} \cr}
$$ 
which are referred to as semi-short \fadho. The conditions \semis{a}\ and
\semis{b}\ are equivalent to the descendant states 
$\bQ_2 |\De,r,j,\bj\rangle^{\rm h.w.}$ and $Q_1 |\De,r,j,\bj\rangle^{\rm h.w.}$
being annihilated by $\bQ_1,{\bar S}^1$ and $Q_2,S^2$ respectively.
Chiral/anti-chiral short multiplets correspond to the BPS conditions
\eqn\quars{\eqalign{
& {\bar t}= 1: \Delta = {\ts{3\over 2}}r \, , \qquad \ \
{\bar Q}_{\dal} |\De,r,j,0\rangle^{\rm h.w.}  =  0 \, ,  \cr
& t = 1: \Delta = -{\ts{3\over 2}}r \, , \qquad 
Q_\alpha |\De,r,0,\bj\rangle^{\rm h.w.}  = 0 \, .\cr}
}
Only if there  are BPS conditions requiring both $t,{\bar t}$ non zero is 
$r$ and hence $\Delta$ fixed and 
the associated supermultiplet is therefore protected.

When $t,{\bar t} = \half$, $Q_2, \,{\bar Q}_1$ are omitted from \vermam, if 
$t,{\bar t}=1$ then $Q_\alpha$, ${\bar Q}_\dal$ are removed respectively. 
As a consequence of $\{Q_\al,{\bar Q}_\dal\}=2P_{\al\dal}$ then for $t,{\bar t}$ 
both non zero particular $P_{\al\dal}$ should be removed from \vermam, thus for 
$t={\bar t}=\half$ $P_{21}$ is dropped. The corresponding Verma module is
denoted by $\V^{t,{\bar t}}_{(\De,r,j,\bj)}$. 

The Verma modules do not form a basis of physical states for a unitary representation,
since in particular the action of $J_-,{\bar J}_-$ in \vermam\ is truncated to ensure
positivity of the norm. A space with positive norm $\H^{t,{\bar t}}_{(\De,r,j,\bj)}$ 
is constructed from the quotient of corresponding Verma module by zero norm 
sub-modules if $2j,2\bj= 0,1,2, \dots$.  
For unitary representations we also require $\Delta \ge 2+2\bj+{\ts{3\over 2}}r,
2+2j-{\ts{3\over 2}}r$ unless one of the BPS conditions in \quars\ hold and 
accordingly then $\Delta = {\ts{3\over 2}}r $ or $- {\ts{3\over 2}}r$. As described 
in \refs{\char,\mep} the characters corresponding to unitary representations are 
constructed from the formal Verma module characters by symmetrising under the Weyl 
group for the maximal compact subgroup of the superconformal group, in this case 
the spin group $SU(2)_J\times SU(2)_\bJ$ with Weyl group $\Bbb Z_2 \times \Bbb Z_2$.

The characters for the Verma modules  $\V^{t,{\bar t}}_{(\De,r,j,\bj)}$
are expressed in terms of variables $s,u,x,\by$
so that in a series expansion of the character the zeroth term is
$s^{2\De}u^{r}x^{2j}\by^{2\bj}$ which  corresponds to the contribution from
the highest weight state. The states in the Verma module in \vermam\
correspond to terms with further factors according to
$ P_{\al\dal} \rightarrow s^2 x^{\pm 1} \by^{\mp 1} $,
$Q_\al \to s \, u^{-1}\, x^{\pm1} $, ${\bar Q}_{\dal} \to s \, u\, \by^{\mp 1}$,
where $\al=1,2$ correspond to $x,x^{-1}$ and $\dal=1,2$ to $\by^{-1},\by$, 
respectively. For $t={\bar t}=0$ the Verma module character, which is written as a 
formal trace, is then
\eqnn\charunit
$$\eqalignno{
\!\!\! C_{(\De,r,j,\bj)}(s,u,x,\by)
&{}= \ttr_{\V_{(\De,r,j,\bj)}}
\big ( s^{2H} \,u^{R}\, x^{2J_3} \, \by^{2{\bar J}_3} \big ) \cr
&{}= s^{2\De} u^{r}C_{j}(x) C_{\bj}(\by)\!\!\!
\sum_{n_{\vep\eta}=0,1,2,\dots,\atop \vep,\eta=\pm 1}\!\!\!
(s^2 x^\vep \by^\eta)^{n_{\vep\eta}}
\!\!\! \sum_{\vep,\eta=\pm 1 \atop n_{\vep},{\bar n}_{\eta}=0,1}\!\!\! 
(s\,u\, x^\vep)^{n_{\vep}}(s\,u^{1}\, \by^\eta)^{{\bar n}_{j\eta}}\cr
&{}= s^{2\De}u^{r}\, C_j(x)C_{\bj}(\by)\,P(s,x,\by)\, \Q(su^{-1},x)\,{\Q}(s u,\by)\,,
& \charunit }
$$
where the factors 
\eqn\defP{
P(s,x,\bx)=\prod_{\vep,\eta=\pm 1}{1\over (1-s^2\, x^\vep \, \bx^\eta)}\,,
\qquad \Q(s,x)=\prod_{\vep=\pm 1}(1+ s\, x^\vep)\, ,
}
arise from the  translation generators and supercharges,
and also
\eqn\vermasut{
C_j(x)=\ttr_{\V_j}\big  ( x^{2J_3}\big ) =\sum_{N=0}^\infty x^{2j-2N}
={x^{2j+2}\over x^2-1}\, ,
}
corresponds to the $SU(2)$ Verma module $\V_j=\{(J_-)^N|j\r^{\rm h.w.}\}$,
$J_3|j\r^{\rm h.w.}=j|j\r^{\rm h.w.}$, $J_+|j\r^{\rm h.w.}=0$.
With shortening conditions the corresponding Verma module character
$C^{t,{\bar t}}_{(\De,r,j,\bj)}(s,u,x,\by)$ has various factors in \charunit\
omitted in accordance with the above discussion.
Since the Weyl group is generated by $x\to x^{-1}, \bx \to \bx^{-1}$ the actual 
characters for physical unitary irreducible representations are then given by
\eqn\fchar{\eqalign{
\chi^{t,{\bar t}}_{(\De,r,j,\bj)}(s,u,x,\bx)={}&\tr_{\H^{t,{\bar t}}_{(\De,r,j,\bj)}}
\big (s^{2H}\,u^{R}\,x^{2J_3}\, \bx^{2\bJ_3} \big )\cr
={}&\sum_{\vep,\eta=\pm 1} C^{t,{\bar t}}_{(\De,r,j,\bj)}(s,u,x^\vep,\by^\eta)\,,}
}
where we may note that
\eqn\sutwochars{
\chi_n(x)= \sum_{\vep=\pm 1}C_{{1\over 2}(n-1)}(x^\vep)
={x^n-x^{-n}\over x-x^{-1}} \, ,
}
is the usual character for the familiar $n$-dimensional $SU(2)$ representation.
For the supertrace in \fchar\ it is sufficient to let $x,\bx \to -x,-\bx$.

For long multiplets all states in the Verma module \vermam\ contribute and
\charunit\ and \fchar\ give,
\eqn\lchar{
\chi^{0,0}_{(\De,r,j,\bj)}(s,u,x,\bx)=s^{2\De}\,u^{r}\,\chi_{2j+1}(x)\,
\chi_{2\bj+1}(\bx)\, P(s,x,\bx)\,\Q(su^{-1},x)\,{\Q}(su,\bx)\,,
}
For semi-short multiplets we have,
\eqn\chirss{\eqalign{
&\chi^{0,{1\over 2}}_{(2\bj+2+{3\over 2}r,r,j,\bj)}(s,u,x,\bx)\cr
&=s^{4\bj+4+3r}\,u^{r}\,\chi_{2j+1}(x)\big(\chi_{2\bj+1}(\bx)+s \,u\,
\chi_{2\bj+2}(\bx) \big) P(s,x,\bx)\,\Q(su^{-1} ,x)\,, \ \ 
r \ge {\ts{2\over 3}}(j-\bj) \, , \cr
&\chi^{{1\over 2},0}_{(2j+2+{3\over 2}r,-r,j,\bj)}(s,u,x,\bx)\cr
&=s^{4j+4+3r}\,u^{-r}\,\big(\chi_{2j+1}(x)+s\, u^{-1}\,\chi_{2j+2}(x)\big)
\chi_{2\bj+1}(\bx) \,P(s,x,\bx)\,{\Q}(s u,\bx) \, , \ \ 
r \ge {\ts{2\over 3}}(\bj-j) \,.}
}
Similarly, for chiral/anti-chiral  short multiplets the superconformal
characters are,
\eqn\bps{\eqalign{
\chi^{0,1}_{{({3\over 2}r,r,j,0)}}(s,u,x,\bx)&{}= (s^{3}u )^{r}\,
\chi_{2j+1}(x)\,P(s,x,\bx)\, \Q(su^{-1} ,x)\,, \ \ r \ge {\ts{2\over 3}}(j+1)\, ,
\cr
\chi^{1,0}_{{({3\over 2}r,-r,0,\bj)}}(s,u,x,\bx)&{}=
(s^{3}u^{-1} )^{r}\,\chi_{2\bj+1}(\bx)\, P(s,x,\bx)\,{\Q}(su,\bx)
\,, \ \ r \ge {\ts{2\over 3}}(\bj+1) \,.}
}
The characters in \bps\
are a special case of those in \chirss\ since
\eqn\spec{\eqalign{
\chi^{0,{1\over 2}}_{(1+{3\over 2}r,r,j,-{1\over 2})}(s,u,x,\bx)
= {}& \chi^{0,1}_{{({3\over 2}(r+1),r+1,j,0)}}(s,u,x,\bx) \, , \cr
\chi^{{1\over 2},0}_{(1+{3\over 2}r,-r,-{1\over 2},\bj)}(s,u,x,\bx)
= {}& \chi^{1,0}_{{({3\over 2}(r+1),-r-1,0,\bj)}}(s,u,x,\bx)\, .}
}

The other cases correspond to protected multiplets. The relevant examples are,
for a  self-conjugate multiplet involving conserved currents,
\eqn\cur{\eqalign{
\chi^{{1\over 2},{1\over 2}}_{{(j+\bj+2,{2\over 3}(j-\bj),j,\bj)}}(s,u,x,\bx)=
u^{{{2\over 3}}(j-\bj)}\big(&{}
\D_{j,\bj}(s,x,\bx)+u^{-1}\,\D_{j+{1\over 2},\bj}(s,x,\bx)\cr
&{}+ u \, \D_{j,\bj+{1\over 2}}(s,x,\bx)+\D_{j+{1\over 2},\bj+{1\over 2}}(s,x,\bx)\big)\,,}
}
where
\eqn\cons{
\D_{j,\bj}(s,x,\bx)=s^{2(j+\bj+2)}\big(\chi_{2j+1}(x)\,\chi_{2\bj+1}(\bx)-s^2
\chi_{2j}(x)\, \chi_{2\bj}(\bx)\big) P(s,x,\bx)\,,
}
is the conformal group character for a $(j,\bj)$ conserved current in four dimensions \mep,
and the Dirac multiplet, with its conjugate, for which the characters are
\eqn\dir{\eqalign{
\chi^{{1\over 2},1}_{{(j+1,{2\over 3}(j+1),j,0)}}(s,u,x,\bx)&{}=u^{{2\over 3}(j+1)}
\big(\E_{j}(s,x,\bx)+ u^{-1}\,\E_{j+{1\over 2}}(s,x,\bx)\big)\,,\cr
\chi^{1,{1\over 2}}_{{(\bj+1,-{2\over 3}(\bj+1),0,\bj)}}(s,u,x,\bx)&{}
=u^{-{2\over 3}(\bj+1)}\big({\overline \E}_{\bj}(s,x,\bx)+u
\,{\overline \E}_{\bj+{1\over 2}}(s,x,\bx)\big)\,, }
}
where
\eqn\free{\eqalign{
\E_{j}(s,x,\bx)&{}=s^{2j+2}\big(\chi_{2j+1}(x)-s^2
\chi_{2j}(x)\,\chi_{2}(\bx)+s^4 \chi_{2j-1}(x)\big)
P(s,x,\bx)\,,\cr
{\overline \E}_{\bj}(s,x,\bx)&{}=s^{2\bj+2}\big(\chi_{2\bj+1}(\bx)-
s^2 \chi_{2}(x)\,\chi_{2\bj}(\bx)+
s^4 \chi_{2\bj-1}(\bx)\big)P(s,x,\bx)\,.}
}
The characters in \dir\ correspond to spin-$j$ chiral/spin-$\bj$ anti-chiral free 
field representations of the conformal group in four dimensions \mep.

At the unitarity threshold the multiplets are reducible which is reflected by
\eqnn\decomp
$$\eqalignno{
\chi^{0,0}_{(2\bj+2+{3\over 2}r,r,j,\bj)}(s,u,x,\bx) = {}& 
\chi^{0,{1\over 2}}_{(2\bj+2+{3\over 2}r,r,j,\bj)}(s,u,x,\bx)\
+ \chi^{0,{1\over 2}}_{(2\bj+{5\over 2}+{3\over 2}r,r+1,j,\bj-{1\over 2})}(s,u,x,\bx)\, , \cr
\chi^{0,{1\over 2}}_{{(j+\bj+2,{2\over 3}(j-\bj),j,\bj)}}(s,u,x,\bx)= {}&
\chi^{{1\over 2},{1\over 2}}_{{(j+\bj+2,{2\over 3}(j-\bj),j,\bj)}}(s,u,x,\bx) \cr
&{}+
\chi^{{1\over 2},0}_{{(j+\bj+{5\over 2},{2\over 3}(j-\bj)-1,j-{1\over 2},\bj)}}(s,u,x,\bx) \, ,
& \decomp }
$$
where we may use \spec\ if $j$ or $\bj$ are zero.

The results for the index in section 2 are equivalent to setting
$1+su\by=0$ and then letting $s\to 0 $ for fixed $t=s^3 u$ and $x$. From \dir\ 
we obtain
\eqn\chind{\eqalign{
\chi^{{1\over 2},1}_{{(j+1,{2\over 3}(j+1),j,0)}}(s,u,-x,-su) \big |_{t=s^3 u}
={}&  (-1)^{2j} \, t^{{2\over 3}(j+1)} \, {\chi_{2j+1}(x) - t \, \chi_{2j}(x) \over
(1-tx)(1-tx^{-1})} \, ,\cr 
\chi^{1,{1\over 2}}_{{(j+1,-{2\over 3}(j+1),0,j)}}(s,u,-x,-su) \big |_{t=s^3 u}
\toinf{s\to 0} {}&  - (-1)^{2j} \, {t^{{4\over 3}(j+1)}  
\over (1-tx)(1-tx^{-1})} \, .}
}
The expressions \indc\ and \indv\  correspond just to the sum of the 
chiral/anti-chiral contributions in \chind\ for $j=0$ and $j=\half$ respectively.

For other characters the limit in \chind\ gives just the following non zero
results 
\eqn\index{\eqalign{
\chi^{0,{1\over 2}}_{(2\bj+2+{3\over 2}r,r,j,\bj)}(s,u,-x,-su)
\big |_{t=s^3 u} \toinf{s\to 0} {}& -(-1)^{2j+2\bj} \
t^{2\bj + 2 + r} {\chi_{2j+1}({x}) \over (1-tx)(1-tx^{-1})} \, , \cr
\chi^{0,1}_{({3\over 2}r,r,j,0)}(s,u,-x,-su) 
\big |_{t=s^3 u} \toinf{s\to 0} {}& (-1)^{2j} \
t^{r}  {\chi_{2j+1}({x}) \over (1-tx)(1-tx^{-1})} \, , \cr
\chi^{{1\over 2},{1\over 2}}_{{(j+\bj+2,{2\over 3}(j-\bj),j,\bj)}} (s,u,-x,-su)
\big |_{t=s^3 u} \toinf{s\to 0} {}& -(-1)^{2j+2\bj} \
t^{{2\over 3}(j+2\bj+3)} {\chi_{2j+1}({x}) \over (1-tx)(1-tx^{-1})} \, .}
}
The expressions in \index\ are relevant for disentangling contributions of
different operators in the expansion of the index in \expI.

\appendix{B}{Characters for Unitary, Symplectic and Orthogonal Groups}

We here give general results for characters for the groups discussed in the text
and verify orthogonality properties in the case of $SU(N)$.

For $SU(n)$ the characters, depending on $\x=(x_1,\dots , x_n)$ subject to 
$\prod_{i=1}^n x_i=1$, are the well known Schur polynomials,
\eqn\schur{
s_{{\underline \lambda}}(\x) = {s}_{(\lambda_1,\dots, \lambda_n)}(\x)=
{{\det}\left[x_i{}^{\lambda_j+n-j}\right]
\over {\det}\left[x_i{}^{n-j}\right]}\, ,
}
where we require ${\underline \lambda}$ to be ordered so that 
$\lambda_1 \ge \lambda_2 \ge \dots \ge \lambda_n$, and
since, as a consequence of the constraint on $\prod_i x_i$,
${s}_{(\lambda_1,\dots, \lambda_n)}(\x) = {s}_{(\lambda_1+c,\dots, \lambda_n+c)}(\x)$
we may also impose $\lambda_n=0$.
In terms of \expforch, $\chi_{SU(n),f}(\x)=s_{(1,0,\dots,0)}(\x)$, 
$\chi_{SU(n), {\bar f}}(\x)=s_{(1,\dots,1,0)}(\x)$ and
$\chi_{SU(n),\rm adj.}(\x)=s_{(2,1,\dots,1,0)}(\x)$. For the Vandermonde determinant
in \van,
\eqn\Van{
\Delta (\x) = {\det}\left[x_i{}^{n-j}\right] \, .
}
As a consistency check we may verify orthogonality of Schur polynomials
$s_{{\underline \lambda}}(\x), s_{{\underline \lambda'}}(\x)$, where both 
${\underline \lambda}, {\underline \lambda'}$ are ordered,
with respect to the measure \invG
\eqn\orth{\eqalign{
& \int_{SU(n)} \!\!\!\!\! \d \mu(\x)\,
s_{{\underline \lambda}}(\x) \, s_{{\underline \lambda'}}(\x) = 
{1\over n!}\int  \prod_{i=1}^{n-1}
{\d x_i \over 2\pi i x_i} \, \Delta(\x)\Delta(\x^{-1}) \, 
s_{{\underline \lambda}}(\x) \, s_{{\underline \lambda'}}(\x)
\bigg |_{\prod_{i=1}^{n} x_i =1 }\cr
& = \int \bigg ( \prod_{i=1}^{n-1} {\d x_i\over 2\pi i x_i}  \, x_i{\!}^{-\lambda_i-n+i} \bigg )
\sum_{\sigma \in \S_n} \hbox{sign}(\sigma) \prod_{j=1}^n (\sigma x_j)^{\lambda'{\!}_j+n-j}
\bigg |_{\prod_{i=1}^{n} x_i =1 }
= \de_{{\underline \lambda},{\underline \lambda'}} \, ,
}}
where the sum is over $n!$ permutations $\sigma$, $\sigma x_j = x_{j^\sigma}$,
belonging to $\S_n$ the Weyl
group for $SU(n)$. The only non zero term surviving the integration in \orth\
is then for $\sigma = e$, the identity, and only when 
${\underline \lambda} = {\underline \lambda'}$.

The Weyl characters for $Sp(2n)$ are also given by the determinantal formula,
\eqn\cspn{
{\widetilde s}_{(\lambda_1,\dots, \lambda_n)}(\x)=
{\det\left[x_i{}^{\lambda_j+n-j+1}-x_i{}^{-\lambda_j-n+j-1}\right]
\over \det\left[x_i{}^{n-j+1}-x_i{}^{-n+j-1}\right] }\, ,
}
with $\lambda_1 \ge \lambda_2 \ge \dots \ge \lambda_n\ge0$. The results in \weylsp\
correspond to ${\chi}_{Sp(2n),f}(\x)={\widetilde s}_{(1,0,\dots, 0)}(\x)$,
${\chi}_{Sp(2n),{\rm adj.}}(\x)={\widetilde s}_{(2,0,\dots, 0)}(\x)$. For the
denominator in \cspn
\eqn\den{
\det\left[x_i{}^{n-j+1}-x_i{}^{-n+j-1}\right]=
\De(\x+\x^{-1})\prod_{i=1}^n(x_i-x_i{}^{-1})\, .
}

For $N=2n$ the characters for $SO(N)$ are given by
\eqn\sotwen{
{\hat s}_{\underline \lambda}(\x)= {{\det}\big [x_i{}^{\lambda_j+n-j}+x_i{}^{-\lambda_j-n+j}\big ]
+ {\det}\big [x_i{}^{\lambda_j+n-j}-x_i{}^{-\lambda_j-n+j}\big ] \over 2\, \De(\x+\x^{-1})}\, ,
}
where $\lambda_1 \ge \lambda_2 \ge \dots \ge |\lambda_n|\ge0$ and
$\chi_{SO(2n),{\rm adj.}}(\x)= {\hat s}_{(1,1,0,\dots,0)}(\x)$.
For $N=2n+1$,
\eqn\sotwon{
{\bar s}_{\underline \lambda}(\x)= {{\det}\big [x_i{}^{\lambda_j+{1\over 2}+n-j}
+x_i{}^{-\lambda_j-{1\over 2}-n+j} \big ] \over 
\De(\x+\x^{-1})\, \prod_{i=1}^n(x_i{}^{1\over 2}-x_i{}^{-{1\over 2}})}\, .
}
where $\lambda_1 \ge \lambda_2 \ge \dots \ge \lambda_n\ge0$ and 
$\chi_{SO(2n+1),{\rm adj.}}(\x)= {\bar s}_{(1,1,0,\dots,0)}(\x)$.

\appendix{C}{Finite $N$ Corrections}

In section 4 we discussed the leading large $N$ expressions for the index, here
we discuss the form of the leading corrections which involve contributions
from operators with non zero baryon number. The expansion of the integral
defining the index generates power symmetric polynomials $p_{\ua}(\z)$
in $\z=(z_1,z_2,\dots)$ as defined in \powersym. We follow a method described in 
\FN\ which relates them to the symmetric Schur polynomials, 
as defined in \schur,
\eqn\schurd{
s_{{\underline \lambda}}(\z) \, , \quad \hbox{where} \quad
\lambda_1 \ge \lambda_2 \ge \dots \ge
\lambda_{\ell({\underline \lambda})} \ge 1 \, , \ \ 
\lambda_{\ell({\underline \lambda})+1 }  = 0 \, .
}
The Schur polynomials are
characters of $SU(N)$ when $\z$ has $N$ components and
$\ell({\underline \lambda}) \le N$. In this case also
\eqn\rest{
{\ts \prod}_{i=1}^N z_i = 1 \quad \Rightarrow \quad s_{{\underline \lambda}}(\z)
= s_{{\underline \lambda}+{\underline \rho}{}_N }(\z) \, ,  \quad
{\underline \rho}{}_N = ( 1, 1, \dots, 1) \, , \ 
\ell({\underline \rho}{}_N) = N \, .
}
The power and Schur symmetric  polynomials are related by
\eqn\broth{
p_{\ua}(\z) =\sum_{{\underline \lambda}\atop \ell({\underline \lambda})\leq N}
\omega_{\ua}{}^{{\underline \lambda}}\, s_{{\underline \lambda}}(\z)\, ,
\qquad |\ua | = |{\underline \lambda}| = {\ts {\sum_n}} \lambda_n  \, .
}
The coefficients $\omega_{\ua}{}^{{\underline \lambda}}$ 
are characters for the symmetric group and they satisfy the completeness relations
\eqn\usefulsec{
\sum_{\underline \lambda}
\omega_{\ua}{}^{{\underline \lambda}}\ \omega_{\ub}{}^{\,{\underline \lambda}}
= z_{\ua}\, \de_{\ua,\ub}\, ,  
}
for $z_{\ua}$ as in \defztau, and \broth\ can be inverted giving
\eqn\invps{
s_{{\underline \lambda}}(\z) = \sum_{\ua} {1\over z_\ua} \, 
\omega_{\ua}{}^{{\underline \lambda}}\ p_{\ua}(\z) \, .
}

The orthogonality relation \orth\
can be extended to, as a consequence of \rest, 
\eqn\serend{
\int_{SU(N)} \!\!\!\!\! \d \mu(\z)\, 
s_{{\underline \lambda}}(\z) \, s_{{\underline \lambda'}}(\z) \bigg 
|_{\ell({\underline \lambda}),\ell({\underline \lambda}') \le N}
= {\de}_{{\underline \lambda}', {\underline \lambda}} + 
{\ts \sum_{n=1}^\infty} \big ( 
\de_{{\underline \lambda}',{\underline \lambda}+ n {\underline \rho}{}_N} + 
\de_{{\underline \lambda}'+ n {\underline \rho}{}_N,{\underline \lambda}}\big )  
\ .
}
Hence
\eqn\frothy{
\int_{SU(N)} \!\!\!\!\!  \d\mu(\z)\, p_\ua(\z)\, p_\ub(\z^{-1})=
\sum_{{\underline \lambda} \atop\ell({\underline \lambda}) \leq N} \Big (
\omega_{\ua}{}^{{\underline \lambda}}\ 
\omega_{\ub}{}^{\,{\underline \lambda}} + {\ts \sum_{n=1}^\infty} \big (
\omega_{\ua}{}^{{\underline \lambda}}\
\omega_{\ub}{}^{\,{\underline \lambda}+ n {\underline \rho}{}_N}
+ \omega_{\ua}{}^{{\underline \lambda}+ n {\underline \rho}{}_N}\, 
\omega_{\ub}{}^{\,{\underline \lambda}} \big ) \Big ) \  .
}

We now consider applying these results to the integral \ithree\ for the
index, where $i(\t,\z)$ is given by \genfor\ but assuming here for simplicity
$h(\t)=f(\t)$ (otherwise there is an additional overall factor as in 
\indexinfNn). Hence the integral becomes
\eqn\neweq{\eqalign{
\I(\t)&=\int_{SU(N)} \!\!\!\!\! \d\mu(\z)\, 
\exp\bigg(\sum_{n=1}^\infty {1\over n}\Big(f(\t^n)p_N(\z^n)p_{N}(\z^{-n})
+g(\t^n)p_N(\z^n)+{\bar g}(\t^n)p_{N}(\z^{-n} )\Big)\bigg)\cr
&=\sum_{\ua,\ub,{\bar \ub}}{1\over z_\ua \, z_\ub\, z_{\bar \ub}} \
f_\ua(\t)\, g_\ub(\t)\, {\bar g}_{\bar \ub}(\t)
\int_{SU(N)}\!\!\!\!\!  \d\mu(\z)\
p_{\ua+\ub}(\z)\, p_{\smash{\ua+{\bar \ub}}}(\z^{-1})\, ,}
}
with the definitions \defztau\ and also
\eqn\infullex{
f_\ua (\t)=\prod_{n\geq 1}f(\t^n)^{a_n}\, , \quad 
g_{\,\ub}(\t)=\prod_{n\geq 1}g(\t^n)^{b_n}\, , \quad 
{\bar g}_{\,{\bar \ub}}(\t)=\prod_{n\geq 1}{\bar g}(\t^n)^{{\bar b}_n}\, .
}
The integral in \frothy\ ensures
\eqn\indf{\eqalign{
\I(\t) = {}& \sum_{\ua,\ub,{\bar \ub}}{1\over z_\ua \, z_\ub\, z_{\bar \ub}} \
f_\ua(\t)\, g_\ub(\t)\, {\bar g}_{\smash{\,{\bar \ub}}}(\t) \cr
\noalign{\vskip -4pt}
& \, {} \times \!\! \sum_{{\underline \lambda} \atop\ell({\underline \lambda}) \leq N} \!\!\!\!
\Big ( \omega_{\ua+\ub}{}^{\,{\underline \lambda}}\
\omega_{\smash{\ua+{\bar\ub}}}{}^{\,{\underline \lambda}} 
+ {\ts \sum_{n=1}^\infty} 
\big ( \omega_{\ua+\ub}{}^{\, {\underline \lambda}}\
\omega_{\smash{\ua+{\bar\ub}}}{}^{\,{\underline \lambda}+ n {\underline \rho}{}_N}
+ \omega_{\ua+\ub}{}^{\,{\underline \lambda}+ n {\underline \rho}{}_N}\,
\omega_{\smash{\ua+{\bar\ub}}}{}^{\,{\underline \lambda}} \big ) \Big ) \, .}
}

Using the completeness relation \usefulsec\ 
the leading term in \indf\ gives, essentially as in \indexinfN,
\eqn\newsqn{
\I_0 (\t)=\sum_{\ua,\ub}{z_{\ua+\ub}\over z_\ua\,  z_\ub{}^2} \, 
f_\ua(\t)\,g_\ub(\t)\, {\bar g}_{\ub}(\t)=
\exp\bigg(\sum_{n=1}^\infty{1\over n}{g(\t^n)\, {\bar g}(\t^n)\over 1-f(\t^n)}\bigg)
\prod_{n=1}^\infty {1\over 1-f(\t^n)}\, .
}
The result \indf\ then shows that
\eqn\defFG{
\I(\t)=\I_0 (\t)+ \I_1(\t)\, ,
}
where
\eqn\newpqn{\eqalign{
\I_1(\t)&{}  =  \sum_{\ua,\ub,{\bar \ub}}{1\over z_\ua \, z_\ub\, z_{\bar \ub}} \
f_\ua(\t)\, g_\ub(\t)\, {\bar g}_{\smash{\,{\bar \ub}}}(\t) \cr
\noalign{\vskip -4pt}
& {} \times \bigg ( \! \sum_{{\underline \lambda} 
\atop\ell({\underline \lambda}) \leq N} \!\!\! \Big (
{\ts \sum_{n=1}^\infty }
\big ( \omega_{\ua+\ub}{}^{\, {\underline \lambda}}\
\omega_{\smash{\ua+{\bar\ub}}}{}^{\,{\underline \lambda}+ n {\underline \rho}{}_N}
+ \omega_{\ua+\ub}{}^{\,{\underline \lambda}+ n {\underline \rho}{}_N}\,
\omega_{\smash{\ua+{\bar\ub}}}{}^{\,{\underline \lambda}} \big ) \Big ) 
- \sum_{{\underline \lambda} \atop\ell({\underline \lambda}) > N} \!\!\!
\omega_{\ua+\ub}{}^{\,{\underline \lambda}}\
\omega_{\smash{\ua+{\bar\ub}}}{}^{\,{\underline \lambda}} \bigg ) \, .}
}
Any sub-leading terms for large $N$ may then be extracted from the 
expression \newpqn\ for $\I_1(\t)$. The first non zero term arises
for $n=1$ and ${\underline \lambda}={\underline 0}$ when \newpqn\ reduces to
\eqn\defGN{
\I_1(\t) \sim \sum_{\ub}{1\over z_\ub}
\big(g_\ub (\t)+{\bar g}_\ub(\t)\big) \,
\omega_{\ub}{}^{\, {\underline \rho}{}_N}\ \, , \qquad |\ub | = N \, .
}

We consider here the application of \defGN\ to the Seiberg and Kutasov-Schwimmer
dual theories, extending the discussion in section 4. Thus we take $N=N_c$
and $N={\tilde N}_c$ and use the leading results for 
$g_{E}(t x,t x^{-1},v,\y,{\tilde \y})$,
${\bar g}_{E}(t x,t x^{-1},v,\y,{\tilde \y})$, which are proportional to 
$t^r$, from \kfghel\ and also $g_{M}(t x,t x^{-1},v,\y,{\tilde \y})$ and 
${\bar g}_{M}(t x,t x^{-1},v,\y,{\tilde \y})$, which are proportional to $t^{2s-r}$, 
from \kfghm. This gives
\eqn\leadingterm{\eqalign{
{I}_{E,1}(tx,t x^{-1},v,\y,{\tilde \y})&\sim
t^{N_c r}\sum_{\ub}{1\over z_\ub}\big(v^{\, N_c} p_\ub(\y)+v^{-N_c}\,
p_\ub({\tilde \y}^{-1})\big)\, 
\omega_{\ub}{}^{\, {\underline \rho}{}_{N_c}}\  \, ,\cr
{I}_{M,1}(tx,t x^{-1},v,\y,{\tilde \y})&\sim
t^{{\tilde N_c}(2s-r)}\sum_{\ub }{1\over z_\ub}\big({\tilde v}^{\,{\tilde N_c}} 
p_\ub(\y^{-1})+ {\tilde v}^{-{\tilde N_c}}\,
p_\ub({\tilde \y})\big)\, 
\omega_{\ub}{}^{\, {\underline \rho}_{\smash {\tilde N}_c}}\ \, ,}
}
with $p_\ub(\y)$ defined as in \infullex.

For Seiberg dual theories then $k=1, \, s = \half$ and from \defN, \defal\
and \vv\ the results in \leadingterm\ are proportional to
$t^{N_c {\tilde N}_c/N_f} v^{N_c}$ for both electric and magnetic cases.
The dependence on $\y,\ty$ is also compatible using \invps
\eqn\eqy{
\sum_{\ub}{1\over z_\ub} \,  p_\ub(\y) \ 
\omega_{\ub}{}^{\, {\underline \rho}{}_{N_c}} = s_{(1^{N_c})}(\y)
= \sum_{\ub}{1\over z_\ub} \, p_\ub(\y^{-1}) \
\omega_{\ub}{}^{{\underline \rho}_{\smash{\tilde N}_c}} 
= s_{(1^{{\tilde N}_c})}(\y^{-1}) \, ,
}
assuming $\prod_i y_i = 1$ and $N_c, {\tilde N}_c \le N_f$.

For Kutasov-Schwimmer dual theories $k=2,3, \dots $ and ${\tilde N}_c$
is as in \defNk\ and $r,s$ are given by \defbega. For this case
\eqn\NNK{
(k+1)N_f\big (  N_c r - {\tilde N_c}(2s-r) \big ) = (k-1)N_f (
k N_f - 2N_c)  \, , \quad k N_f - 2N_c = 2 {\tilde N}_c  - k N_f \, .
}
In consequence the powers of $t$ in \leadingterm\ do not match.
If $k N_f - 2 N_c <0$ then we must have $N_c > N_f$ and then
\eqn\rewt{
\sum_{\ub}{1\over z_\ub} \,  p_\ub(\y) \
\omega_{\ub}{}^{\, {\underline \rho}{}_{N_c}} = 0 \, , 
}
so that the leading contribution to ${I}_{E,1}(tx,t x^{-1},v,\y,{\tilde \y})$
in \leadingterm\ 
vanishes. Conversely if $k N_f - 2 N_c > 0$ we must have ${\tilde N}_c > N_f$
and the leading contribution to 
${I}_{M,1}(tx,t x^{-1},v,\y,{\tilde \y})$ is absent. 
In consequence there is no manifest inconsistency between $I_E$ and $I_M$
beyond the large $N$ limit and so perhaps further
evidence for matching of the index for the electric
and magnetic dual Kutasov-Schwimmer theories.

\appendix{D}{Useful Identities}

We here note some useful properties of the elliptic Gamma functions and other
infinite products defined by \eGam\ and \defth. As well as \mGam\ we may also define
\eqn\eprod{\eqalign{
(x_1,\dots,x_n;q) = {}& (x_1;q)\cdots (x_n;q) \, ,\cr
\theta(x_1,\dots,x_n;q) = {}&  \theta(x_1;q)\cdots \theta(x_n;q) \, ,}
}
for $(x;q),\theta(x;q)$ in \defth. Useful identities are
\eqn\fgtr{ (x;q) = (x;q^2)\, (xq;q^2)\,,\qquad
(x;q^2)=({\sqrt{x}};q) \, (-{\sqrt{x}};q)\, ,
}
which extend also to $\theta(x;q)$. For the latter we may also note
\eqn\thid{
\theta (qx;q) = \theta(x^{-1};q) = - {1\over x}\, \theta(x;q) \, .
}
In terms of standard Jacobi theta functions $(q^2,q^2) \, \theta (q e^{2iu}, q^2)
= \vartheta_4(u,q)$. The Jacobi product identity is equivalent to
\eqn\Jid{
(q;q) \,\theta(x;q) = \sum_{n=-\infty}^\infty (-1)^n q^{{1\over 2}n(n-1)} x^n \,,
}
while the addition formula in the form
\eqn\addth{
a \, \theta( ba,ba^{-1},cz,cz^{-1};p) + b\, \theta( cb,cb^{-1},ca,ca^{-1};p)
+ c \,\theta( ac,ac^{-1},bz,bz^{-1};p) = 0 \, ,
}
with notation as in \eprod, is significant later.

For the elliptic gamma function,
properties which prove useful are, besides the reflection formula \GG,
\eqn\propelg{
\Gamma(xq;p,q)=\theta(x;p)\,\Gamma(x;p,q)\,,\qquad 
\Gamma(xp;p,q)=\theta(x;q)\,\Gamma(x;p,q)\,,
}
and 
\eqn\ido{
\Gamma(p;p,q)=(q;q)/(p;p)\, , \qquad
\Gamma(q;p,q)=(p;p)/(q;q)\,,
}
and
\eqn\idtw{
\Gamma(-1;p,q)={1\over 2(-q;q)\, (-p;p)}\,,
}
so that, using also \fgtr,
\eqn\idth{
\Gamma(-1;p,q)\,\Gamma(-p;p,q)=\half {(p,p^2)^2}\, , \qquad
\Gamma(-1;p,q)\,\Gamma(-q;p,q)=\half {(q,q^2)^2}\, .
}
With the notation in \mGam\ we have
\eqn\GthP{
\Gamma(z,z^{-1};p,q) = {1\over \theta(z;q)\, \theta(z^{-1};p)} \, ,
}
which may be rewritten in various forms with the aid of \thid.

\appendix{E}{Verification of the Spiridonov Elliptic Beta Integral}

We here describe an approach to showing  $\A(p,q,\hu)=\B(p,q,\hu)$, as defined
in \IA\ and  \indexm, analogous to that outlined for
the Nassrallah-Rahman theorem in section 5. From its definition in
\defl\ and appendix D, with the notation in \eprod, we have that
\eqn\varchth{
\hI(p,q,qu_1,u_2,\dots,z)=
{\theta(u_1 z,u_1/z;p)\over \theta(\lambda z,\lambda/z;p)}\, \hI(p,q,\hu,z)\,,
}
so that using the identity, which follows from \addth\ and \thid,
\eqn\thidr{
u_2\theta(u_1 z,u_1/z,\lambda u_2,\lambda/u_2;p)- 
u_1\theta(u_2 z, u_2/z, \lambda u_1,\lambda/u_1;p)
=- u_1 \theta (u_1u_2,u_2/u_1, \lambda z,\lambda/z;p)\,,
}
(for $p=0$ this reduces to \fourtid)
we find that $\hI(p,q,\hu,z)$ satisfies the $q$-difference relation
\eqn\qdifft{\eqalign{
&u_2\theta (\lambda u_2,\lambda/u_2;p)\, \hI(p,q,qu_1,u_2,\dots,z)
-u_1\theta(\lambda u_1,\lambda/u_1;p)\, \hI(p,q,u_1,qu_2,\dots,z)\cr
&=-u_1\theta(u_1 u_2,u_2/u_1;p)\, \hI(p,q,\hu,z)\,.}
}
Since this holds for any $z$ the $q$-difference relation extends to $\A(p,q,\hu)$.

Similarly,
\eqn\varchth{
\B(p,q,qu_1,u_2,\dots,u_5)=
\prod_{a=1}^5{\theta(u_1u_a;p)\over \theta(\lambda/u_a;p)}\, \B(p,q,\hu)\,,
}
so that using the  identity, which is also equivalent to \addth,
\eqn\thrupo{\eqalign{
& u_2 \theta(u_1u_3,u_1u_4,u_1u_5,\lambda u_2;p)
-u_1 \theta(u_2u_3,u_2u_4,u_2u_5,\lambda u_1;p)\cr
& {} =- u_1\theta(u_2/u_1,\lambda/u_3,\lambda/u_4,\lambda/u_5;p)\,,}
}
assuming $\lambda$ as in \defl,
it is easy to show that $\B(p,q,\hu,z)$ also satisfies \qdifft.

The proof is now essentially the same as that described in section 5. 
$\A(p,q,\hu)$, $\B(p,q,\hu)$ are both are analytic functions in each $u_a$ so it is
sufficient to show that they are equal for a particular non zero choice of $\hu$
and use the $q$-difference relation to extend this to an infinite discrete set of
$\hu$ which then, by analyticity, implies $\A(p,q,\hu) =\B(p,q,\hu)$ for arbitrary 
$\hu$  so long as both are non singular.

We then consider the same special case 
chosen for proving the Nassrallah-Rahman theorem,
$\hu_0=(u,1,-1,q^{1\over 2},-q^{1\over 2})$.
For $\hI$ as in \defl\ we have 
\eqn\idio{
\hI(p,q,\hu_0,z)=
-z^2\, {1\over \theta (z^2;q)\, \theta(z^2;p^2)\, \theta (u z, u /z;p)}\,,
}
and in \IA,
\eqn\intA{
\A(p,q,\hu_0)  
= (p;p)\, (q;q) \, {1\over 4 \pi i}\oint{\d z\over z} \, 
{(z^2 p ;p^2) \, (z^{-2}p;p^2) \over \theta (u z, u /z;p)} \, .
}
Using \ido, \idtw\ and \idth, we may show, for $\B$ as in \indexm, that
\eqn\intB{
\B(p,q,\hu_0)={(q;q)\over (p;p)}\, {1\over 2\theta (u^2;p^2)}\,.
}
Thus to show equality of \intA\ and \intB\ it is necessary to verify that
\eqn\whatshow{
\F(u,p) = {1\over 2 \pi i}\oint {\d z\over z}\,  \I_0(u,z,p)
={1\over (p;p)^2 }\, {1\over \theta(u^2;p^2)}\,, \quad 
\I_0(u,z,p) = { \theta (z^2 p ;p^2)\over \theta (u z, u /z;p)} \, ,
}
where, requiring  $p<|u|<1$, the $z$-integration is around the unit circle.

Spiridonov \spi\ evaluated the integral in \whatshow\ by using rather non trivial 
identities. We here present a simpler argument. The integrand $\I_0(u,z,p)$
has poles inside the contour $|z|=1$ at $z=up^n, p^{n+1}/u$ and outside
at $z=p^{-n}/u, u p^{-n-1}$, for $n=0,1,2,\dots$, and satisfies, from \thid,
\eqn\Izid{
\I_0(pu,z,p) = u^2 \I_0(u,z,p) \, .
}
If we let $u\to p u$ \Izid\ would naively imply that a similar relation 
holds for $\F(u,p)$ but under this change the pole at $z=p/u$ moves outside
the contour while the one at $z=u/p$ moves inside. Taking into account
the contributions of these poles we get
\eqn\Fdiff{
\F(pu,p) = u^2 \F(u,p) + {2\over (p;p)^2 }\, {1\over \theta(u^2p^2;p^2)}
=  u^2 \bigg ( \F(u,p) -  {2\over (p;p)^2 }\, {1\over \theta(u^2;p^2)}\bigg ) \, .
}
The form of the integral in \whatshow\ shows that $\F(u,p)$  has poles solely at 
$u^2 = p^r$ for some positive or negative integer $r$, then \Fdiff\ and analyticity 
implies that $\F(u,p)$ can only have the form given by the result in \whatshow.

\listrefs

\bye